\documentclass[12pt]{article}

\input{epsf}
\usepackage{rotating,here,epsfig}
\usepackage{cite}
\newcommand{\T}        {\textstyle}

\newcommand{\sz}{{S}_0(\Omega)}

\newcommand{\dsig}{\frac{d \sigma}{d \Omega}}
\newcommand{\oi}{{\cal O}_i}

\newcommand{\e}        {\mathrm{e}}
\newcommand{\E}        {\mathrm{E}}
\newcommand{\M}        {\mathrm{M}}
\newcommand{\W}        {\rm W}

\newcommand{\ee}       {\mbox{$e^+ e^-$}}
\newcommand{\ww}       {\mbox{$ {\mathrm W}^+ {\mathrm W}^-$}}
\newcommand{\zz}       {\mbox{$ {\mathrm Z} {\mathrm Z}$}}
\newcommand{\wwz}      {\mbox{$ \mathrm{WWZ}$}}
\newcommand{\wwgamma}  {\mbox{$ \mathrm{WW}\gamma$}}

\newcommand{\eeww}     {\mbox{$ \ee \rightarrow \ww$}}
\newcommand{\eew}     {\mbox{$ \ee \rightarrow  \mathrm{W} e \nu$}}
\newcommand{\eeg}     {\mbox{$ \ee \rightarrow  \nu \overline{\nu} \gamma (\gamma)$}}
\newcommand{\lvlv}     {\mbox{$ \ell \nu \ell \nu$}}
\newcommand{\lvqq}     {\mbox{$ \ell \nu q \bar{q}$}}
\newcommand{\evqq}     {\mbox{$ e \nu q \bar{q}$}}
\newcommand{\muvqq}     {\mbox{$ \mu \nu q \bar{q}$}}

\newcommand{\tauvqq}     {\mbox{$ \tau \nu q \bar{q}$}}

\newcommand{\evev}     {\mbox{$ e\nu e\nu $}}
\newcommand{\evmuv}     {\mbox{$ e\nu \mu\nu $}}
\newcommand{\evtauv}     {\mbox{$ e\nu \tau\nu $}}

\newcommand{\qqqq}     {\mbox{$ q \bar{q} q \bar{q}$}}

\newcommand{\mev}      {\mbox{MeV}}
\newcommand{\gev}      {\mbox{GeV}}
\newcommand{\gevm}      {\mbox{GeV}/\mathrm{c}^2}

\newcommand{\pbinv}    {\mbox{pb$^{-1}$}}

\newcommand{\kg}{\kappa_\gamma}

\newcommand{\lmg}{\lambda_\gamma}

\newcommand{\dkg}{\mbox{$\Delta\kappa_\gamma$}}
\newcommand{\lgam}{\mbox{$\lambda_\gamma$}}
\newcommand{\dgz}{\mbox{$\Delta{\mathrm g}^{\mathrm Z}_1$}}
\newcommand{\myrule}{\rule[-4.mm]{0mm}{9mm} }

\def\qqp     { {\mathrm{q}}{\bar{\mathrm{q}}'}}

\newcommand{\xe}{\mbox{$x_{E}$}}
\newcommand{\tg}{\mbox{$\theta_{\gamma}$}}
\newcommand{\eg}{\mbox{$E_{\gamma}$}}
\newcommand{\egz}{\mbox{$E_{\gamma}^{\mathrm{Z}}$}}
\newcommand{\sqs}{\mbox{$\sqrt s$}}

\def\ap#1#2#3   {{Ann. Phys. (NY)} {#1} (#2) #3}
\def\apj#1#2#3  {{Astrophys. J.} {#1} (#2) #3}
\def\apjl#1#2#3 {{Astrophys. J. Lett.} {#1} (#2) #3}
\def\app#1#2#3  {{Acta. Phys. Pol.} {#1} (#2) #3}
\def\ar#1#2#3   {{Ann. Rev. Nucl. Part. Sci.} {#1} (#2) #3}
\def\cpc#1#2#3  {{Comp. Phys. Commun.} {#1} (#2) #3}
\def\epj#1#2#3  {{Eur. Phys. J.} {#1} (#2) #3}
\def\err#1#2#3  {{Erratum} {#1} (#2) #3}
\def\ib#1#2#3   {{ibid.} {#1} (#2) #3}
\def\jmp#1#2#3  {{J. Math. Phys.} {#1} (#2) #3}
\def\ijmp#1#2#3 {{Int. J. Mod. Phys.} {#1} (#2) #3}
\def\jetp#1#2#3 {{JETP Lett.} {#1} (#2) #3}
\def\jpg#1#2#3  {{J. Phys. G.} {#1} (#2) #3}
\def\mpl#1#2#3  {{Mod. Phys. Lett.} {#1} (#2) #3}
\def\nat#1#2#3  {{Nature (London)} {#1} (#2) #3}
\def\nc#1#2#3   {{Nuovo Cim.} {#1} (#2) #3}
\def\nim#1#2#3  {{Nucl. Instrum. and Methods} {#1} (#2) #3}
\def\np#1#2#3   {{Nucl. Phys.} {#1} (#2) #3}
\def\pcps#1#2#3 {{Proc. Cam. Phil. Soc.} {#1} (#2) #3}
\def\pl#1#2#3   {{Phys. Lett.} {#1} (#2) #3}
\def\prep#1#2#3 {{Phys. Rep.} {#1} (#2) #3}
\def\prev#1#2#3 {{Phys. Rev.} {#1} (#2) #3}
\def\prl#1#2#3  {{Phys. Rev. Lett.} {#1} (#2) #3}
\def\prs#1#2#3  {{Proc. Roy. Soc.} {#1} (#2) #3}
\def\ptp#1#2#3  {{Prog. Th. Phys.} {#1} (#2) #3}
\def\ps#1#2#3   {{Physica Scripta} {#1} (#2) #3}
\def\rmp#1#2#3  {{Rev. Mod. Phys.} {#1} (#2) #3}
\def\rpp#1#2#3  {{Rep. Prog. Phys.} {#1} (#2) #3}
\def\sjnp#1#2#3 {{Sov. J. Nucl. Phys.} {#1} (#2) #3}
\def\spj#1#2#3  {{Sov. Phys. JEPT} {#1} (#2) #3}
\def\spu#1#2#3  {{Sov. Phys.-Usp.} {#1} (#2) #3}
\def\zp#1#2#3   {{Zeit. Phys.} {#1} (#2) #3}

\begin{document}
\begin{titlepage}
\begin{center}
  {\large EUROPEAN ORGANIZATION FOR NUCLEAR RESEARCH (CERN)}
\end{center}
\vspace{1.0 cm}
\begin{flushright}
  CERN-EP/2001-022\\*[5mm]
  February 28, 2001\\*[5mm]
\end{flushright}
\begin{center}
\vspace{2cm}
  {\Large\bf Measurement of Triple Gauge-Boson Couplings\\[3mm] at LEP energies 
up to 189~\gev} \\
  \vspace{1.0cm} \large {\large The ALEPH
Collaboration$^{*)}$
}
\end{center}
\vspace{4.0cm}
\begin{abstract}
  The triple gauge-boson couplings involving the W are determined using
  data samples collected with the ALEPH detector at mean centre-of-mass 
  energies of 183~\gev\ and 189~\gev, corresponding to integrated 
  luminosities of 57~\pbinv\ and 174~\pbinv, respectively.
  The couplings, ${\mathrm g}^{\mathrm Z}_1$, $\kg$\ and $\lmg$, are
  measured using W-pair events, single-W production and
  single-$\gamma$ production. Each coupling is measured individually
  with the other two couplings fixed at their Standard Model 
  value. Including 
  ALEPH results from lower energies, the 95\%\ confidence level intervals
  for the deviation to the Standard Model are
$$\begin{array}{rcccl}
-0.087 < & \dgz  & < &0.141 \\
-0.200 < & \dkg  & < &0.258 \\
-0.062 < & \lgam & < &0.147.
\end{array}$$

Fits are also presented where two or all three couplings are
allowed to vary. In addition, W-pair events are
used to set limits on the C- or P-violating couplings
$\mathrm{g^V_4}$, $\mathrm{g^V_5}$, $\mathrm{\tilde{\kappa}_{V}}$, and
$\mathrm{\tilde{\lambda}_{V}}$, where V denotes either $\gamma$ or Z.
No deviations from the Standard Model expectations are observed.
\end{abstract}
\vspace{0.5cm}
\begin{center}
  ({\it Submitted to European Physical Journal C})
\end{center}
\vfill
\vbox{
\hrule width7.5cm height0.5pt\vskip4pt
\noindent $^*$) See next pages for the list of authors.}

\end{titlepage}
 
\pagestyle{empty}
\newpage
\small
%
%
\newlength{\saveparskip}
\newlength{\savetextheight}
\newlength{\savetopmargin}
\newlength{\savetextwidth}
\newlength{\saveoddsidemargin}
\newlength{\savetopsep}
\setlength{\saveparskip}{\parskip}
\setlength{\savetextheight}{\textheight}
\setlength{\savetopmargin}{\topmargin}
\setlength{\savetextwidth}{\textwidth}
\setlength{\saveoddsidemargin}{\oddsidemargin}
\setlength{\savetopsep}{\topsep}
%
%
\setlength{\parskip}{0.0cm}
\setlength{\textheight}{25.0cm}
\setlength{\topmargin}{-1.5cm}
\setlength{\textwidth}{16 cm}
\setlength{\oddsidemargin}{-0.0cm}
\setlength{\topsep}{1mm}
\pretolerance=10000
\centerline{\large\bf The ALEPH Collaboration}
\footnotesize
\vspace{0.5cm}
{\raggedbottom
\begin{sloppypar}
\samepage\noindent
A.~Heister,
S.~Schael
\nopagebreak
\begin{center}
\parbox{15.5cm}{\sl\samepage
Physikalisches Institut das RWTH-Aachen, D-52056 Aachen, Germany}
\end{center}\end{sloppypar}
\vspace{2mm}
\begin{sloppypar}
\noindent
R.~Barate,
I.~De~Bonis,
D.~Decamp,
P.~Ghez,
C.~Goy,
S.~Jezequel,
J.-P.~Lees,
F.~Martin,
E.~Merle,
\mbox{M.-N.~Minard},
B.~Pietrzyk,
B.~Trocm\'e
\nopagebreak
\begin{center}
\parbox{15.5cm}{\sl\samepage
Laboratoire de Physique des Particules (LAPP), IN$^{2}$P$^{3}$-CNRS,
F-74019 Annecy-le-Vieux Cedex, France}
\end{center}\end{sloppypar}
\vspace{2mm}
\begin{sloppypar}
\noindent
S.~Bravo,
M.P.~Casado,
M.~Chmeissani,
J.M.~Crespo,
E.~Fernandez,
M.~Fernandez-Bosman,
Ll.~Garrido,$^{15}$
E.~Graug\'{e}s,
J.~Lopez,
M.~Martinez,
G.~Merino,
R.~Miquel,
Ll.M.~Mir,
A.~Pacheco,
D.~Paneque,
H.~Ruiz
\nopagebreak
\begin{center}
\parbox{15.5cm}{\sl\samepage
Institut de F\'{i}sica d'Altes Energies, Universitat Aut\`{o}noma
de Barcelona, E-08193 Bellaterra (Barcelona), Spain$^{7}$}
\end{center}\end{sloppypar}
\vspace{2mm}
\begin{sloppypar}
\noindent
A.~Colaleo,
D.~Creanza,
N.~De~Filippis,
M.~de~Palma,
G.~Iaselli,
G.~Maggi,
M.~Maggi,$^{1}$
S.~Nuzzo,
A.~Ranieri,
G.~Raso,$^{24}$
F.~Ruggieri,
G.~Selvaggi,
L.~Silvestris,
P.~Tempesta,
A.~Tricomi,$^{3}$
G.~Zito
\nopagebreak
\begin{center}
\parbox{15.5cm}{\sl\samepage
Dipartimento di Fisica, INFN Sezione di Bari, I-70126 Bari, Italy}
\end{center}\end{sloppypar}
\vspace{2mm}
\begin{sloppypar}
\noindent
X.~Huang,
J.~Lin,
Q. Ouyang,
T.~Wang,
Y.~Xie,
R.~Xu,
S.~Xue,
J.~Zhang,
L.~Zhang,
W.~Zhao
\nopagebreak
\begin{center}
\parbox{15.5cm}{\sl\samepage
Institute of High Energy Physics, Academia Sinica, Beijing, The People's
Republic of China$^{8}$}
\end{center}\end{sloppypar}
\vspace{2mm}
\begin{sloppypar}
\noindent
D.~Abbaneo,
P.~Azzurri,
T.~Barklow,$^{30}$
G.~Boix,$^{6}$
O.~Buchm\"uller,
M.~Cattaneo,
F.~Cerutti,
B.~Clerbaux,
G.~Dissertori,
H.~Drevermann,
R.W.~Forty,
M.~Frank,
F.~Gianotti,
T.C.~Greening,
J.B.~Hansen,
J.~Harvey,
D.E.~Hutchcroft,
P.~Janot,
B.~Jost,
M.~Kado,
V.~Lemaitre,$^{23}$
P.~Maley,
P.~Mato,
A.~Moutoussi,
F.~Ranjard,
L.~Rolandi,
D.~Schlatter,
P.~Spagnolo,
W.~Tejessy,
F.~Teubert,
E.~Tournefier,$^{26}$
A.~Valassi,
J.J.~Ward,
A.E.~Wright
\nopagebreak
\begin{center}
\parbox{15.5cm}{\sl\samepage
European Laboratory for Particle Physics (CERN), CH-1211 Geneva 23,
Switzerland}
\end{center}\end{sloppypar}
\vspace{2mm}
\begin{sloppypar}
\noindent
Z.~Ajaltouni,
F.~Badaud,
S.~Dessagne,
A.~Falvard,$^{20}$
D.~Fayolle,
P.~Gay,
P.~Henrard,
J.~Jousset,
B.~Michel,
S.~Monteil,
\mbox{J-C.~Montret},
D.~Pallin,
J.M.~Pascolo,
P.~Perret,
F.~Podlyski
\nopagebreak
\begin{center}
\parbox{15.5cm}{\sl\samepage
Laboratoire de Physique Corpusculaire, Universit\'e Blaise Pascal,
IN$^{2}$P$^{3}$-CNRS, Clermont-Ferrand, F-63177 Aubi\`{e}re, France}
\end{center}\end{sloppypar}
\vspace{2mm}
\begin{sloppypar}
\noindent
J.D.~Hansen,
J.R.~Hansen,
P.H.~Hansen,
B.S.~Nilsson,
A.~W\"a\"an\"anen
\nopagebreak
\begin{center}
\parbox{15.5cm}{\sl\samepage
Niels Bohr Institute, 2100 Copenhagen, DK-Denmark$^{9}$}
\end{center}\end{sloppypar}
\vspace{2mm}
\begin{sloppypar}
\noindent
G.~Daskalakis,
A.~Kyriakis,
C.~Markou,
E.~Simopoulou,
A.~Vayaki
\nopagebreak
\begin{center}
\parbox{15.5cm}{\sl\samepage
Nuclear Research Center Demokritos (NRCD), GR-15310 Attiki, Greece}
\end{center}\end{sloppypar}
\vspace{2mm}
\begin{sloppypar}
\noindent
A.~Blondel,$^{12}$
\mbox{J.-C.~Brient},
F.~Machefert,
A.~Roug\'{e},
M.~Swynghedauw,
R.~Tanaka
\linebreak
H.~Videau
\nopagebreak
\begin{center}
\parbox{15.5cm}{\sl\samepage
Laboratoire de Physique Nucl\'eaire et des Hautes Energies, Ecole
Polytechnique, IN$^{2}$P$^{3}$-CNRS, \mbox{F-91128} Palaiseau Cedex, France}
\end{center}\end{sloppypar}
\vspace{2mm}
\begin{sloppypar}
\noindent
E.~Focardi,
G.~Parrini,
K.~Zachariadou
\nopagebreak
\begin{center}
\parbox{15.5cm}{\sl\samepage
Dipartimento di Fisica, Universit\`a di Firenze, INFN Sezione di Firenze,
I-50125 Firenze, Italy}
\end{center}\end{sloppypar}
\vspace{2mm}
\begin{sloppypar}
\noindent
A.~Antonelli,
M.~Antonelli,
G.~Bencivenni,
G.~Bologna,$^{4}$
F.~Bossi,
P.~Campana,
G.~Capon,
V.~Chiarella,
P.~Laurelli,
G.~Mannocchi,$^{5}$
F.~Murtas,
G.P.~Murtas,
L.~Passalacqua,
M.~Pepe-Altarelli$^{25}$
\nopagebreak
\begin{center}
\parbox{15.5cm}{\sl\samepage
Laboratori Nazionali dell'INFN (LNF-INFN), I-00044 Frascati, Italy}
\end{center}\end{sloppypar}
\vspace{2mm}
\begin{sloppypar}
\noindent
M.~Chalmers,
A.W.~Halley,
J.~Kennedy,
J.G.~Lynch,
P.~Negus,
V.~O'Shea,
B.~Raeven,
D.~Smith,
A.S.~Thompson
\nopagebreak
\begin{center}
\parbox{15.5cm}{\sl\samepage
Department of Physics and Astronomy, University of Glasgow, Glasgow G12
8QQ,United Kingdom$^{10}$}
\end{center}\end{sloppypar}
\vspace{2mm}
\begin{sloppypar}
\noindent
S.~Wasserbaech
\nopagebreak
\begin{center}
\parbox{15.5cm}{\sl\samepage
Department of Physics, Haverford College, Haverford, PA 19041-1392, U.S.A.}
\end{center}\end{sloppypar}
\vspace{2mm}
\begin{sloppypar}
\noindent
R.~Cavanaugh,
S.~Dhamotharan,
C.~Geweniger,
P.~Hanke,
V.~Hepp,
E.E.~Kluge,
G.~Leibenguth,
A.~Putzer,
K.~Tittel,
S.~Werner,$^{19}$
M.~Wunsch$^{19}$
\nopagebreak
\begin{center}
\parbox{15.5cm}{\sl\samepage
Kirchhoff-Institut f\"ur Physik, Universit\"at Heidelberg, D-69120
Heidelberg, Germany$^{16}$}
\end{center}\end{sloppypar}
\vspace{2mm}
\begin{sloppypar}
\noindent
R.~Beuselinck,
D.M.~Binnie,
W.~Cameron,
G.~Davies,
P.J.~Dornan,
M.~Girone,$^{1}$
N.~Marinelli,
J.~Nowell,
H.~Przysiezniak,$^{2}$
S.~Rutherford,
J.K.~Sedgbeer,
J.C.~Thompson,$^{14}$
R.~White
\nopagebreak
\begin{center}
\parbox{15.5cm}{\sl\samepage
Department of Physics, Imperial College, London SW7 2BZ,
United Kingdom$^{10}$}
\end{center}\end{sloppypar}
\vspace{2mm}
\begin{sloppypar}
\noindent
V.M.~Ghete,
P.~Girtler,
E.~Kneringer,
D.~Kuhn,
G.~Rudolph
\nopagebreak
\begin{center}
\parbox{15.5cm}{\sl\samepage
Institut f\"ur Experimentalphysik, Universit\"at Innsbruck, A-6020
Innsbruck, Austria$^{18}$}
\end{center}\end{sloppypar}
\vspace{2mm}
\begin{sloppypar}
\noindent
E.~Bouhova-Thacker,
C.K.~Bowdery,
D.P.~Clarke,
G.~Ellis,
A.J.~Finch,
F.~Foster,
G.~Hughes,
R.W.L.~Jones,$^{1}$
M.R.~Pearson,
N.A.~Robertson,
M.~Smizanska
\nopagebreak
\begin{center}
\parbox{15.5cm}{\sl\samepage
Department of Physics, University of Lancaster, Lancaster LA1 4YB,
United Kingdom$^{10}$}
\end{center}\end{sloppypar}
\vspace{2mm}
\begin{sloppypar}
\noindent
I.~Giehl,
F.~H\"olldorfer,
K.~Jakobs,
K.~Kleinknecht,
M.~Kr\"ocker,
A.-S.~M\"uller,
H.-A.~N\"urnberger,
G.~Quast,
B.~Renk,
E.~Rohne,
H.-G.~Sander,
S.~Schmeling,
H.~Wachsmuth,
C.~Zeitnitz,
T.~Ziegler
\nopagebreak
\begin{center}
\parbox{15.5cm}{\sl\samepage
Institut f\"ur Physik, Universit\"at Mainz, D-55099 Mainz, Germany$^{16}$}
\end{center}\end{sloppypar}
\vspace{2mm}
\begin{sloppypar}
\noindent
A.~Bonissent,
J.~Carr,
P.~Coyle,
C.~Curtil,
A.~Ealet,
D.~Fouchez,
O.~Leroy,
T.~Kachelhoffer,
P.~Payre,
D.~Rousseau,
A.~Tilquin
\nopagebreak
\begin{center}
\parbox{15.5cm}{\sl\samepage
Centre de Physique des Particules de Marseille, Univ M\'editerran\'ee,
IN$^{2}$P$^{3}$-CNRS, F-13288 Marseille, France}
\end{center}\end{sloppypar}
\vspace{2mm}
\begin{sloppypar}
\noindent
M.~Aleppo,
S.~Gilardoni,
F.~Ragusa
\nopagebreak
\begin{center}
\parbox{15.5cm}{\sl\samepage
Dipartimento di Fisica, Universit\`a di Milano e INFN Sezione di
Milano, I-20133 Milano, Italy.}
\end{center}\end{sloppypar}
\vspace{2mm}
\begin{sloppypar}
\noindent
A.~David,
H.~Dietl,
G.~Ganis,$^{27}$
K.~H\"uttmann,
G.~L\"utjens,
C.~Mannert,
W.~M\"anner,
\mbox{H.-G.~Moser},
R.~Settles,$^{1}$
H.~Stenzel,
G.~Wolf
\nopagebreak
\begin{center}
\parbox{15.5cm}{\sl\samepage
Max-Planck-Institut f\"ur Physik, Werner-Heisenberg-Institut,
D-80805 M\"unchen, Germany\footnotemark[16]}
\end{center}\end{sloppypar}
\vspace{2mm}
\begin{sloppypar}
\noindent
J.~Boucrot,$^{1}$
O.~Callot,
M.~Davier,
L.~Duflot,
\mbox{J.-F.~Grivaz},
Ph.~Heusse,
A.~Jacholkowska,$^{20}$
L.~Serin,
\mbox{J.-J.~Veillet},
I.~Videau,
J.-B.~de~Vivie~de~R\'egie,$^{28}$
C.~Yuan
\nopagebreak
\begin{center}
\parbox{15.5cm}{\sl\samepage
Laboratoire de l'Acc\'el\'erateur Lin\'eaire, Universit\'e de Paris-Sud,
IN$^{2}$P$^{3}$-CNRS, F-91898 Orsay Cedex, France}
\end{center}\end{sloppypar}
\vspace{2mm}
\begin{sloppypar}
\noindent
G.~Bagliesi,
T.~Boccali,
G.~Calderini,
V.~Ciulli,
L.~Fo\`a,
A.~Giammanco,
A.~Giassi,
F.~Ligabue,
A.~Messineo,
F.~Palla,
G.~Sanguinetti,
A.~Sciab\`a,
G.~Sguazzoni,
R.~Tenchini,$^{1}$
A.~Venturi,
P.G.~Verdini
\samepage
\begin{center}
\parbox{15.5cm}{\sl\samepage
Dipartimento di Fisica dell'Universit\`a, INFN Sezione di Pisa,
e Scuola Normale Superiore, I-56010 Pisa, Italy}
\end{center}\end{sloppypar}
\vspace{2mm}
\begin{sloppypar}
\noindent
O.~Awunor,
G.A.~Blair,
J.~Coles,
G.~Cowan,
A.~Garcia-Bellido,
M.G.~Green,
L.T.~Jones,
T.~Medcalf,
A.~Misiejuk,
J.A.~Strong,
P.~Teixeira-Dias
\nopagebreak
\begin{center}
\parbox{15.5cm}{\sl\samepage
Department of Physics, Royal Holloway \& Bedford New College,
University of London, Egham, Surrey TW20 OEX, United Kingdom$^{10}$}
\end{center}\end{sloppypar}
\vspace{2mm}
\begin{sloppypar}
\noindent
R.W.~Clifft,
T.R.~Edgecock,
P.R.~Norton,
I.R.~Tomalin
\nopagebreak
\begin{center}
\parbox{15.5cm}{\sl\samepage
Particle Physics Dept., Rutherford Appleton Laboratory,
Chilton, Didcot, Oxon OX11 OQX, United Kingdom$^{10}$}
\end{center}\end{sloppypar}
\vspace{2mm}
\begin{sloppypar}
\noindent
\mbox{B.~Bloch-Devaux},$^{1}$
D.~Boumediene,
P.~Colas,
B.~Fabbro,
E.~Lan\c{c}on,
\mbox{M.-C.~Lemaire},
E.~Locci,
P.~Perez,
J.~Rander,
\mbox{J.-F.~Renardy},
A.~Rosowsky,
P.~Seager,$^{13}$
A.~Trabelsi,$^{21}$
B.~Tuchming,
B.~Vallage
\nopagebreak
\begin{center}
\parbox{15.5cm}{\sl\samepage
CEA, DAPNIA/Service de Physique des Particules,
CE-Saclay, F-91191 Gif-sur-Yvette Cedex, France$^{17}$}
\end{center}\end{sloppypar}
\vspace{2mm}
\begin{sloppypar}
\noindent
N.~Konstantinidis,
A.M.~Litke,
C.~Loomis,
G.~Taylor
\nopagebreak
\begin{center}
\parbox{15.5cm}{\sl\samepage
Institute for Particle Physics, University of California at
Santa Cruz, Santa Cruz, CA 95064, USA$^{22}$}
\end{center}\end{sloppypar}
\vspace{2mm}
\begin{sloppypar}
\noindent
C.N.~Booth,
S.~Cartwright,
F.~Combley,
P.N.~Hodgson,
M.~Lehto,
L.F.~Thompson
\nopagebreak
\begin{center}
\parbox{15.5cm}{\sl\samepage
Department of Physics, University of Sheffield, Sheffield S3 7RH,
United Kingdom$^{10}$}
\end{center}\end{sloppypar}
\vspace{2mm}
\begin{sloppypar}
\noindent
K.~Affholderbach,
A.~B\"ohrer,
S.~Brandt,
C.~Grupen,
J.~Hess,
A.~Ngac,
G.~Prange,
U.~Sieler
\nopagebreak
\begin{center}
\parbox{15.5cm}{\sl\samepage
Fachbereich Physik, Universit\"at Siegen, D-57068 Siegen, Germany$^{16}$}
\end{center}\end{sloppypar}
\vspace{2mm}
\begin{sloppypar}
\noindent
C.~Borean,
G.~Giannini
\nopagebreak
\begin{center}
\parbox{15.5cm}{\sl\samepage
Dipartimento di Fisica, Universit\`a di Trieste e INFN Sezione di Trieste,
I-34127 Trieste, Italy}
\end{center}\end{sloppypar}
\vspace{2mm}
\begin{sloppypar}
\noindent
H.~He,
J.~Putz,
J.~Rothberg
\nopagebreak
\begin{center}
\parbox{15.5cm}{\sl\samepage
Experimental Elementary Particle Physics, University of Washington, Seattle,
WA 98195 U.S.A.}
\end{center}\end{sloppypar}
\vspace{2mm}
\begin{sloppypar}
\noindent
S.R.~Armstrong,
K.~Cranmer,
P.~Elmer,
D.P.S.~Ferguson,
Y.~Gao,$^{29}$
S.~Gonz\'{a}lez,
O.J.~Hayes,
H.~Hu,
S.~Jin,
J.~Kile,
P.A.~McNamara III,
J.~Nielsen,
W.~Orejudos,
Y.B.~Pan,
Y.~Saadi,
I.J.~Scott,
\mbox{J.H.~von~Wimmersperg-Toeller}, 
J.~Walsh,
W.~Wiedenmann,
J.~Wu,
Sau~Lan~Wu,
X.~Wu,
G.~Zobernig
\nopagebreak
\begin{center}
\parbox{15.5cm}{\sl\samepage
Department of Physics, University of Wisconsin, Madison, WI 53706,
USA$^{11}$}
\end{center}\end{sloppypar}
}
\footnotetext[1]{Also at CERN, 1211 Geneva 23, Switzerland.}
\footnotetext[2]{Now at LAPP, 74019 Annecy-le-Vieux, France}
\footnotetext[3]{Also at Dipartimento di Fisica di Catania and INFN Sezione di
 Catania, 95129 Catania, Italy.}
\footnotetext[4]{Deceased.}
\footnotetext[5]{Also Istituto di Cosmo-Geofisica del C.N.R., Torino,
Italy.}
\footnotetext[6]{Supported by the Commission of the European Communities,
contract ERBFMBICT982894.}
\footnotetext[7]{Supported by CICYT, Spain.}
\footnotetext[8]{Supported by the National Science Foundation of China.}
\footnotetext[9]{Supported by the Danish Natural Science Research Council.}
\footnotetext[10]{Supported by the UK Particle Physics and Astronomy Research
Council.}
\footnotetext[11]{Supported by the US Department of Energy, grant
DE-FG0295-ER40896.}
\footnotetext[12]{Now at Departement de Physique Corpusculaire, Universit\'e de
Gen\`eve, 1211 Gen\`eve 4, Switzerland.}
\footnotetext[13]{Supported by the Commission of the European Communities,
contract ERBFMBICT982874.}
\footnotetext[14]{Also at Rutherford Appleton Laboratory, Chilton, Didcot, UK.}
\footnotetext[15]{Permanent address: Universitat de Barcelona, 08208 Barcelona,
Spain.}
\footnotetext[16]{Supported by the Bundesministerium f\"ur Bildung,
Wissenschaft, Forschung und Technologie, Germany.}
\footnotetext[17]{Supported by the Direction des Sciences de la
Mati\`ere, C.E.A.}
\footnotetext[18]{Supported by the Austrian Ministry for Science and Transport.}
\footnotetext[19]{Now at SAP AG, 69185 Walldorf, Germany}
\footnotetext[20]{Now at Groupe d' Astroparticules de Montpellier, Universit\'e de Montpellier II, 34095 Montpellier, France.}
\footnotetext[21]{Now at D\'epartement de Physique, Facult\'e des Sciences de Tunis, 1060 Le Belv\'ed\`ere, Tunisia.}
\footnotetext[22]{Supported by the US Department of Energy,
grant DE-FG03-92ER40689.}
\footnotetext[23]{Now at Institut de Physique Nucl\'eaire, D\'epartement de Physique, Universit\'e Catholique de Louvain, 1348 Louvain-la-Neuve, Belgium.}
\footnotetext[24]{Also at Dipartimento di Fisica e Tecnologie Relative, Universit\`a di Palermo, Palermo, Italy.}
\footnotetext[25]{Now at CERN, 1211 Geneva 23, Switzerland.}
\footnotetext[26]{Now at ISN, Institut des Sciences Nucl\'eaires, 53 Av. des Martyrs, 38026 Grenoble, France.}
\footnotetext[27]{Now at INFN Sezione di Roma II, Dipartimento di Fisica, Universit\`a di Roma Tor Vergata, 00133 Roma, Italy.}
\footnotetext[28]{Now at Centre de Physique des Particules de Marseille, Univ M\'editerran\'ee, F-13288 Marseille, France.}
\footnotetext[29]{Also at Department of Physics, Tsinghua University, Beijing, The People's Republic of China.}
\footnotetext[30]{Also at SLAC, Stanford, CA 94309, U.S.A.}
\setlength{\parskip}{\saveparskip}
\setlength{\textheight}{\savetextheight}
\setlength{\topmargin}{\savetopmargin}
\setlength{\textwidth}{\savetextwidth}
\setlength{\oddsidemargin}{\saveoddsidemargin}
\setlength{\topsep}{\savetopsep}
\normalsize
\newpage
\pagestyle{plain}
\setcounter{page}{1}

\pagestyle{plain}
\setcounter{page}{1}
\setcounter{footnote}{0}
 

\section{Introduction}
The existence of the triple gauge-boson couplings (TGC) is a direct 
consequence of the $\mbox{SU(2)}_{\mathrm{L}} \times \mbox{U(1)}_{\mathrm{Y}}$ 
gauge theory. The measurement of the TGCs represents a fundamental
test of the non-Abelian nature of the Standard Model. The
triple \wwgamma\  and \wwz\  couplings have been studied at LEP in
\ee\ collisions at energies above the \mbox{W-pair} production threshold,
using direct W-pair production (\eeww)~\cite{ref-wwtgc-172,ref-lepwwtgc},
single-W production (\eew) and single-$\gamma$ production
(\eeg)~\cite{ref-singlew-183,ref-singleg-183,ref-lepevW}. Measurements
of the TGCs have also been made at the Tevatron from studies of di-boson
production~\cite{ref-tgc-TEV}. This paper presents new results for the
TGCs from analyses of W-pair, single-W, and single-$\gamma$ final
states using data recorded in 1997 and 1998 with the ALEPH detector. In
1997 and 1998 ALEPH recorded total integrated luminosities of 
56.81~\pbinv\ and 174.20~\pbinv, at mean centre-of-mass energies of
182.66~\gev\ and 188.63~\gev, denoted as 183 and 189~\gev. \par

The most general Lorentz invariant parametrisation of the \wwgamma\
and \wwz\ vertices can be described by 14 independent complex
couplings~\cite{ref-hagivara,ref-bilenky,ref-lep2report}, 7 for each
vertex: $\mathrm{g^V_1}$, $\mathrm{g^V_4}$, 
$\mathrm{g^V_5}$, $\mathrm{\kappa_{V}}$, $\mathrm{\lambda_{V}}$,
$\mathrm{\tilde{\kappa}_{V}}$ and $\mathrm{\tilde{\lambda}_{V}}$,
where V denotes either $\gamma$ or Z. Assuming electromagnetic gauge
invariance, C- and P-conservation, the set of 14 couplings can be
reduced to 5 parameters: $\mathrm{g^Z_1}$, $\kappa_{\gamma}$,
$\mathrm{\kappa_{Z}}$, $\lambda_{\gamma}$ and $\mathrm{\lambda_{Z}}$,
with Standard Model values $\mathrm{g^Z_1 = \kappa_{Z} =
  \kappa_{\gamma} = 1}$ and $\mathrm{\lambda_{Z} = 
  \lambda_{\gamma} = 0}$.  
Precision measurements at the Z resonance at LEP and SLC
also provide bounds on the
couplings~\cite{ref-lowlep,ref-low}. However, local
$\mathrm{SU(2)_{L}\times U(1)_{Y}}$ gauge invariance reduces the
relevance of these bounds~\cite{ref-lowlep} and introduces the
constraints:
$$\mathrm{ \Delta \kappa_{Z} = - \Delta
  \kappa_{\gamma}\tan^2\theta_{w} + \Delta g_1^Z },$$
$$\mathrm{\lambda_{Z} = \lambda_{\gamma}},$$
where $\Delta$ denotes the deviation
of the respective quantity from its non-zero Standard Model value, and
$\theta_{\mathrm W}$ is the weak mixing angle. Hence, only three
parameters remain: \dgz, \dkg, and \lgam~\cite{ref-lep2report}.

Using data from \eeww\ final states all three 
couplings \dgz, \dkg\ and \lgam\ can be tested, whereas the
single-W and single-$\gamma$ final states allow measurements of only
the \mbox{\wwgamma-couplings}, \dkg\ and \lgam.
Although the contribution from W-pair production  dominates the
combined  limits, the single-W and single-$\gamma$ events provide
complementary information, which enhances the sensitivity
especially for \dkg.\par

In this analysis the three couplings \dgz, \dkg\ and \lgam\ are measured 
individually with the two other couplings fixed at zero, their 
Standard Model value. Fits are also presented, where two or all three 
couplings are allowed to vary.\par

The C- or  P-violating sector of the TGCs is weakly bound. Indirect
limits on $\mathrm{\tilde{\kappa}_{\gamma}}$,
$\mathrm{\tilde{\lambda}_{\gamma}}$, $\mathrm{\tilde{\kappa}_{Z}}$ and
$\mathrm{g^Z_4}$ exist, while there are no direct or indirect limits on the
parameters $\mathrm{\tilde{\lambda}_{Z}}$, $\mathrm{g^\gamma_4}$,
$\mathrm{g^\gamma_5}$ and $\mathrm{g^Z_5}$~\cite{ref-low}. Only the
parameter $\mathrm{\tilde{\lambda}_{\gamma}}$ is tightly constrained
by precision low-energy measurements~\cite{ref-lmtglow}. This paper
includes, for the first time, single-parameter fits to the
unconstrained real and imaginary parts of the 8 couplings 
$\mathrm{g^V_4}$,  $\mathrm{g^V_5}$, $\mathrm{\tilde{\kappa}_{V}}$,
and $\mathrm{\tilde{\lambda}_{V}}$, all zero in the Standard Model,
based on an analysis of semileptonic (\evqq\ and \muvqq) W-pair
events.\par

The paper is organised as follows. In Section~\ref{sec-aleph}, a brief
description of the ALEPH detector is given. The Monte Carlo event
generators used in the analyses are 
presented in Section~\ref{sec-samples}. The analysis of the single-$\gamma$
final states is described in Section~\ref{sec-singleg} and
Section~\ref{sec-singlew} is devoted to the single-W analysis. The
description of the two analyses is rather concise, as they are
presented in earlier
publications~\cite{ref-singlew-183,ref-singleg-183}. In
Section~\ref{sec-ww} the measurement of TGCs from W-pair events is
discussed in detail. Finally, all measurements are combined with ALEPH
results from \ww\ production at 172~\gev~\cite{ref-wwtgc-172},
single-W production at 183~\gev~\cite{ref-singlew-183} and
single-$\gamma$ production at 183~\gev~\cite{ref-singleg-183}. The resulting
single- and multi-parameter fits are discussed in Section~\ref{sec-combined}, 
followed by a summary and conclusions in Section~\ref{sec-conclusion}. 
\section{The ALEPH Detector}
\label{sec-aleph}
A detailed description of the ALEPH detector can be
found in Ref.~\cite{ref-aleph,ref-perf}.
The central part of the ALEPH detector is dedicated to the
reconstruction of the trajectories of charged particles. Following a
charged particle from the interaction point outwards, the trajectory
is measured by a two-layer silicon strip vertex detector (VDET), a
cylindrical drift chamber and a large time projection chamber (TPC).
The three tracking detectors are immersed in a $1.5$~T axial field
provided by a superconducting solenoidal coil. Altogether they
measure charged particle momenta with a resolution of $\delta
  p_{\mathrm T} /p_{\mathrm T} = 6\times 10^{-4}p_{\mathrm T} \oplus
0.005$ ($p_{\mathrm T}$ in
\gev$\mathrm{/c})$. In the following, charged particle tracks
reconstructed with at least one hit in the VDET, at least four hits in
the TPC, and originating from within a cylinder of length $20$ cm and
$2$ cm radius centred on the nominal interaction point and parallel
with the beam, are referred to as {\em good tracks}.\par

Photons and electrons are identified in the electromagnetic calorimeter 
(ECAL), situated between the TPC and the coil. It is a 
lead--proportional-wire sampling calorimeter segmented in $0.9^{\circ}
\times 0.9^{\circ}$ towers read out in three sections in depth. It has a
total thickness of 22 radiation lengths and yields a relative energy
resolution of $0.18/\sqrt{E}+0.009$, with $E$ in \gev, for isolated
photons. At low polar
angles, the ECAL is supplemented by two calorimeters, LCAL and SiCAL,
principally used to measure the integrated luminosity collected by the
experiment. Electrons are identified by their
transverse  and longitudinal shower profiles in ECAL and their
specific ionisation in the  TPC. A detailed description of the photon
identification can be  found in~\cite{ref-perf}.\par

The iron return yoke is equipped with 23 layers of streamer tubes and
forms the hadron calorimeter (HCAL). The latter provides a
relative energy resolution of charged and neutral hadrons of
$0.85/\sqrt{E}$, with $E$ in \gev . Muons are distinguished from
hadrons by their distinct pattern in HCAL and by the muon chambers
composed of two double-layers of streamer tubes outside HCAL.\par

The information from the tracking detectors and the calorimeters are
combined in an energy flow algorithm~\cite{ref-perf}. For each event,
the algorithm provides a set of charged and neutral reconstructed
particles, called {\em energy flow objects}, which are used in the
analysis. Studies of $\mathrm{Z\rightarrow q\overline{q}}$ events show
that the angular resolution of jets reconstructed from energy flow
objects is typically 30~mrad in space and the energy resolution
approximately $\sigma_E = (0.6\sqrt{E}+0.6) (1+\cos^2
\theta)$~\gev, where $E$ is the jet energy in \gev\ and $\theta$ is
the polar angle with respect to the z-axis along the e$^-$ beam
direction.
\section{Monte Carlo generators}
\label{sec-samples}
Samples of fully simulated events, reconstructed with the same
program as the data, are used for the design of the selections,
determination of the signal efficiencies and the 
estimation of the background. The size of the generated signal samples
correspond to 20 times (for the single-$\gamma$ and single-W analysis)
and up to 80 times (for the W-pair analysis) the collected
luminosity.\par

The efficiency for the single-$\gamma$ cross section measurement is
estimated using a modified version of the {\tt
KORALZ}~\cite{ref-koralz} Monte Carlo program. The {\tt KORALZ}
generator simulates initial state photons using
YFS~exponentiation~\cite{ref-YFS}. The generator is modified to
include the effects from photons produced as bremsstrahlung off the
exchanged virtual W. This treatment includes the expected Standard
Model contribution and possible anomalous couplings together with the
interference of the two. The effect on the overall cross section
is found to be small
($\sim$0.2\%) for Standard Model couplings. However, it can be as
large as a few percent in certain
kinematical regions. The predictions of the modified {\tt KORALZ}
Monte Carlo are confirmed by an independent generator {\tt
NUNUGPV}~\cite{ref-nunugpv}, which is based on exact lowest order
amplitudes for the production of up to three photons in the final
state, modified for higher order QED effects using transverse momentum
dependent structure functions.\par

For the single-W study the GRC4F program~\cite{ref-grc4f} is used 
to simulate the four-fermion signal process final state $e \nu
f\bar{f}$.  The effective QED coupling constant is fixed
to be $\alpha_{\mathrm{QED}} = 1/130.2$ as suggested
in~\cite{ref-hagiwara2}. For initial state radiation, the photon
structure function approach is utilised. Final state radiation
and tau decays are simulated with {\tt PHOTOS}~\cite{ref-photos} and
{\tt TAUOLA}~\cite{ref-tauola}.\par

For the analysis of W-pair final states, the {\tt
KORALW}~\cite{ref-koralw} generator, which includes all
four-fermion diagrams contributing to \ww -like final states, is used
to produce the primary reference  sample with a W mass of
80.35~\gev$\mathrm{/c^2}$. The {\tt KORALW} generator is interfaced
with {\tt JETSET}~\cite{ref-pythia}, {\tt PHOTOS}~\cite{ref-photos},
and {\tt TAUOLA}~\cite{ref-tauola} for fragmentation, final state
radiation and $\tau$ decays, respectively. In addition, several
samples are generated using the double resonant
CC03~\cite{ref-lep2report} diagrams with non-standard values for one
coupling at a time, to check the reconstruction and TGC
determination. Finally, a sample generated with the double resonant
CC03 diagrams is used to optimise selection efficiencies and
parametrise the corrections used in the kinematic fitting.\par

In order to include the effects from various background processes,
Monte Carlo samples are generated with a corresponding
integrated luminosity of each background sample of at least 20 times
that of the data. {\tt PYTHIA}~\cite{ref-pythia} is used to generate 
\ee$\rightarrow q\bar{q}(\gamma)$, \zz,  $\mathrm{Zee}$,
and \ee$\rightarrow \mathrm{We\nu}$ event samples. In the \zz\ sample,
events with \ww -like final states are discarded to avoid double
counting. Two-photon processes
are simulated with the
{\tt PHOT02}~\cite{ref-phot02} generator. The {\tt
KORALZ}~\cite{ref-koralz} and {\tt UNIBAB}~\cite{ref-unibab}
generators are used for the di-lepton final states.
\section{Single-$\gamma$ production analysis}
\label{sec-singleg}
Events with one or more photons and missing energy can be used to
probe the anomalous \wwgamma\ coupling parameters \dkg\ and
\lgam. Although the single-$\gamma$ channel is less sensitive to the
couplings compared to the W-pair and single-W channels, it provides
complementary information. A detailed description of the Standard
Model processes involved in the reaction \eeg\ and the modelling of
the measured triple gauge-boson couplings can be found in
\cite{ref-singleg-183}.\par

The sensitivity to the \wwgamma\ couplings in the single-$\gamma$
channel comes from the W-W fusion diagram. The W's exchanged in this
t-channel diagram are predominantly at low momentum transfer.
The single-$\gamma$ channel is therefore mainly sensitive to \dkg\ because
contributions from \lgam\ contain higher powers of the W
momenta. Furthermore, the effect of anomalous TGCs depends on the energy
of the photon. For low energy photons, below the radiative return to
the Z peak the sensitivity arises from the interference between the
Standard Model and the anomalous contribution; this interference has a
linear dependence on the TGCs. In the region around the radiative
return to
the Z peak, the sensitivity is minimal. For high energy photons, above
the radiative return to the Z peak, the dependence on the TGCs is
quadratic.
\subsection{Event selection and determination of the TGCs}
The events are selected from the 189~\gev\ data sample using the
procedure described in
\cite{ref-multig-183}. In summary, single-$\gamma$ events are selected
by requiring at least one photon candidate with $\tg > 20^{\circ}$
and $p_{\mathrm{T}}^{\gamma}/E_{\mathrm{beam}}>0.1$ and no additional
activity in
form of reconstructed charged tracks or energy deposits in the forward
regions (below 14$^{\circ}$). Events where a photon has converted into a
$\ee$ pair are not considered.

\begin{figure}[H]
\begin{center}
\begin{picture}(350,500)
\put(-60,-10){\mbox{\epsfig{file=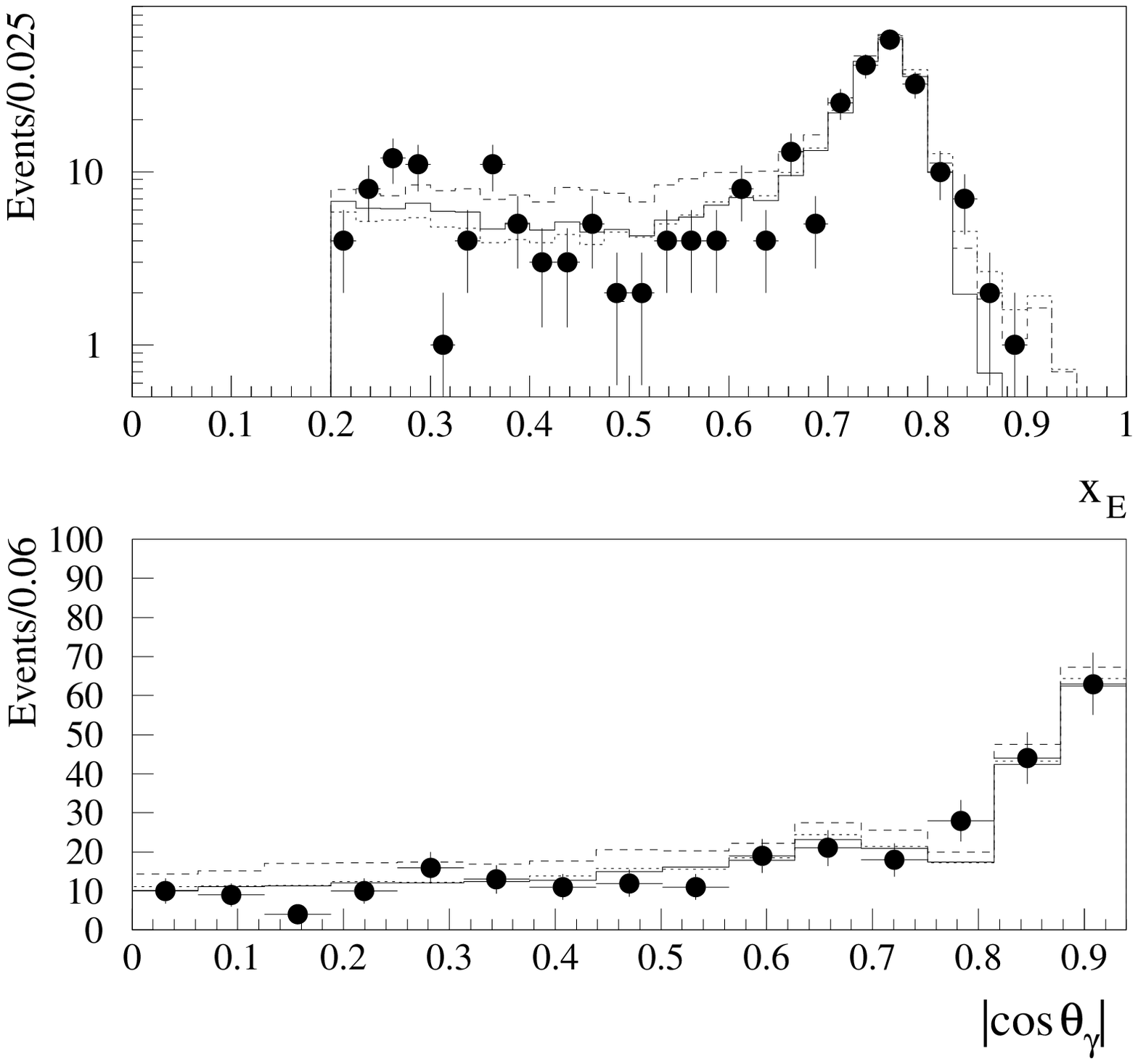,scale=1.}}}
\put(140,447){\bf{\Huge ALEPH}}
\put(25,385){\Large a)}
\put(25,175){\Large b)}
\end{picture}
\end{center}
\caption{Distribution of a) the scaled energy, $x_E$, and
b) the absolute value of the cosine of the polar angle for single-$\gamma$ events selected in
189~\gev\ data. The data are represented by solid dots, while the
solid histogram shows the distribution for the Standard Model. The dashed
and dotted histograms show the distribution for non-standard values of $\dkg=\pm 5.0$.}
\label{fig-singleg_cos_xe}
\end{figure}

Anomalous contributions to the \wwgamma\ vertex increase the total
cross section and lead to characteristic energy and angular distributions
of the final state photons. For the single-$\gamma$ channel the TGCs
are extracted from the data by performing a
maximum likelihood fit based on the overall number of
observed photons, their polar angles $\theta_{\gamma}$ and scaled
energies $\xe (=E_{\gamma}/E_{beam})$ of the form
\begin{equation}
\log L = \log\frac{(N_{\mathrm{exp}})^{N_{\mathrm{obs}}} \,
                  \mathrm{e}^{-N_{\mathrm{exp}}}}  {N_{\mathrm{obs}} !}
        +\sum \log P_i,
\label{eq-lldef}
\end{equation}
where $P_{i}$ is the probability density function of observing event
$i$ with a given value of $\xe$ and $\tg$ and $N_{\mathrm{exp}}$ is the
expected number of events including background. The probability
density function and the expected number of events for different
values of the couplings are constructed by reweighting fully simulated
single-$\gamma$ events. Distributions of the polar angle and the
scaled energy for single-$\gamma$ events are shown in
Figure~\ref{fig-singleg_cos_xe}.

Two separate kinematic regions are
used in the fit, excluding a region around the Z peak return, where
the sensitivity is small. Defining $ \egz =
(s-m^2_{\mathrm{Z}})/(2\sqs )$, the excluded region is $\egz -
3\Gamma_{\mathrm{Z}} < \eg < \egz + 0.5\Gamma_{\mathrm{Z}}$.
The total numbers of photons used in the fit are 120 (128 expected)
and 260 (258 expected) below and above the excluded region.

\subsection{Results}
At present energies, the cross section and the shape contribute
equally in the likelihood function for \dkg, whereas the result for
\lgam\ is dominated by the sensitivity to the shape above the excluded
region. The
estimation of the systematic uncertainties follows the procedure
described in Ref.~\cite{ref-singleg-183} and the different contributions
are summarised in Table~\ref{tab-singleg_sys}.
\begin{table}
\caption{\footnotesize Summary of the systematic errors on single
  parameter fits for
\dkg\ and \lgam\ from the single-$\gamma$ analysis at 189~\gev.}
\begin{center}
\begin{tabular}{|l|c|c|} \hline
Source & \dkg & \lgam \\ \hline \hline
Acceptance corrections & $0.08$ & $0.08$ \\
Photon energy calibration & $0.11$ & $0.14$ \\
Background & $0.05$ & $0.05$ \\
Luminosity & $0.03$ & $0.03$ \\
Theoretical uncertainty & $0.13$ & $0.15$ \\ \hline
Total & $0.20$ & $0.22$ \\ \hline
\end{tabular}
\end{center}
\label{tab-singleg_sys}
\end{table}
The fitted results for the 189~\gev\ data for
single parameter fits, where each coupling is
determined setting the other coupling to its Standard Model value, are
\begin{eqnarray*}
\dkg &=&0.4\pm{0.7}\pm{0.2} \qquad (\lgam = 0)  \\ 
\lgam&=&0.3\pm{0.9}\pm{0.2} \qquad (\dkg = 0)
\end{eqnarray*}
where the first error is the statistical error and the second
is the systematic uncertainty. The 95\%
confidence level limits including systematic errors are:
\begin{eqnarray*}
       -1.1 <& \dkg &< 1.8    \qquad (\lgam = 0)\\
       -1.5 <& \lgam &< 2.0      \qquad (\dkg=0).
\end{eqnarray*}
The validity of these 95\% C.L. limits and the error from the
likelihood fit have been checked using many Monte Carlo samples
corresponding to the data luminosity as described in
\cite{ref-singleg-183}.

\begin{figure}[H]
\begin{center}
\begin{picture}(350,300)
\put(-75,-210){\mbox{\epsfig{file=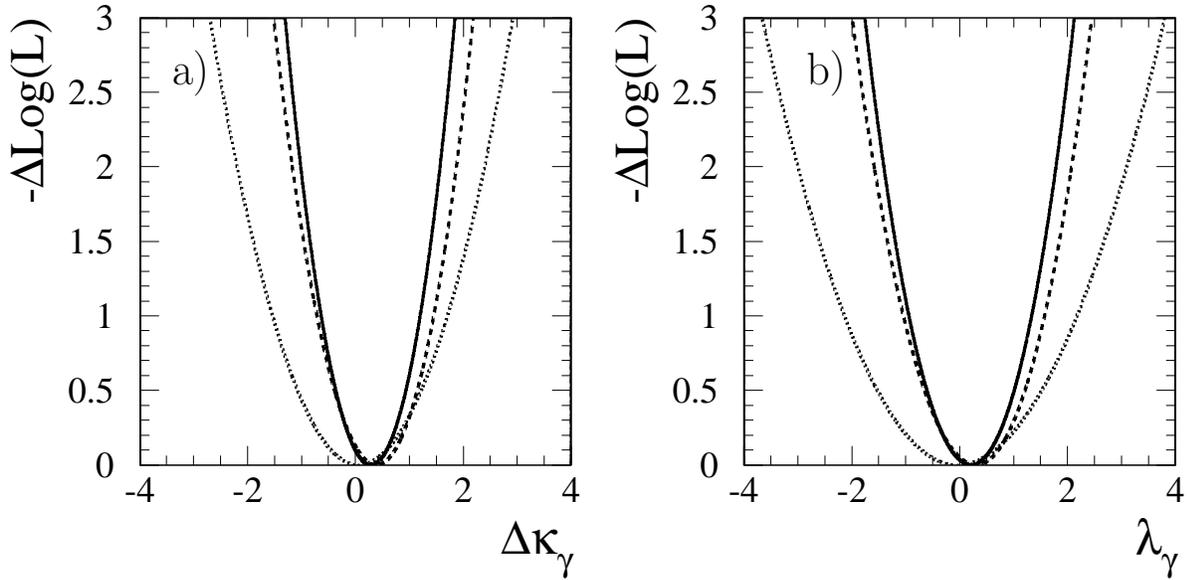,scale=0.85}}}
\put(140,247){\bf{\Huge ALEPH}}
\put(5,200){\Large a)}
\put(245,200){\Large b)}
\end{picture}
\end{center}
\caption{Negative log-likelihood curves, including systematic
uncertainties, from the single-$\gamma$
analysis for a) $\dkg$ and b) $\lgam$ for the 189~\gev\ data (dashed
line), 161 - 183~\gev\ data~\cite{ref-singleg-183} (dotted line), and
the combined results (solid line). The curve for each coupling is
obtained while fixing the other coupling to its Standard Model
value.
}
\label{f:likeli2}
\end{figure}

Combining with the previous measurement for centre-of-mass energies
between 161 and 183~GeV~\cite{ref-singleg-183}, the 95\% C.L. limits
on \dkg\ and \lgam\ from single-$\gamma$ production are
\begin{displaymath}
   \begin{array}{rcccrlcr}
     -1.0 & < & \dkg & < & 1.5 & \ (\lgam  & = & 0), \\
     -1.4 & < & \lgam  & < & 1.8 & \ (\dkg & = & 0).
   \end{array}
\end{displaymath}
The negative log-likelihood functions curves are shown in
Figure~\ref{f:likeli2}
for the 189~\gev\ data, 161 - 183\,GeV data~\cite{ref-singleg-183}
and the combined results. In the combination, the systematic errors
from acceptance and theoretical prediction are assumed to be fully
correlated, while all other sources are taken as uncorrelated.
\section{Single-W production analysis}
\label{sec-singlew}
Single-W production, \eew, is sensitive to the 
\wwgamma\ vertex. This sensitivity comes from the $\gamma$-W fusion
diagram, where the momentum transfer is low. As for
the single-$\gamma$ channel, the single-W channel is therefore mostly
sensitive to \dkg~\cite{ref-tsukamoto}.
\subsection{Selection}
The analysis of single-W production is performed on the 189~\gev\ data
sample. All W decay modes are used and the selection of each W decay final state, described in the
following, has been optimised for the single-W signal definition used
in the previous analysis at lower centre-of-mass
energies~\cite{ref-singlew-183}:
\begin{displaymath}
   \left\{
        \begin{array}{ll}
        \theta_{\e} < 34\,\mbox{mrad}, &  \\ 
        \E_{\ell} > 20\,\gev\mbox{ and } |\cos\theta_{\ell}|<0.95 
        &  \mbox{for leptonic decays}, \\
        \M_{\qqp} > 60\,\gevm &  \mbox{for hadronic decays},
        \end{array}
   \right.      
\end{displaymath}
where $\theta_\e$ is the polar angle of the scattered electron,
$\E_{\ell}$ and $\theta_{\ell}$ are the energy and polar angle of
leptons from the W decay. $\M_{\qqp}$ is the invariant
mass of the quark pair. The cut angle at 34 mrad corresponds to the
lower edge of the acceptance of the ALEPH detector.

As single-W
production is dominated by t-channel processes, the
outgoing electron is predominantly emitted at small polar angles.
Another specific feature is the large missing momentum carried away by 
the electron-neutrino, and therefore a common selection criterium for
all single-W final states is the requirement of the missing momentum
direction to be within the detector acceptance,
$|\cos\theta_{\mathrm{miss}}|<0.9$.
\subsubsection{Leptonic Selection}
The leptonic W decay is characterised by a high energy isolated
lepton. Allowing for a multi-prong decay of the tau, events with one
or three good tracks ($|\cos\theta|<0.95$) are accepted. The
selection cuts are the same as in the analysis in Ref.~\cite{ref-singlew-183} and are summarised here.\par

In addition to the cut on the missing momentum direction, tagged
two-photon events are rejected by requiring that no energy be detected
within a cone of $12^\circ$ around the beam axis (E$_{12}=0$).\par

The remaining backgrounds, mainly untagged two-photon events and 
two-fermion events, are eliminated by requiring that the transverse 
missing momentum be greater than 0.06$\sqrt{s}$.
This threshold is increased to 0.1$\sqrt{s}$ if the missing momentum
direction points to within 10$^\circ$ in azimuth to the boundaries 
between the two LCAL halves or between the six inner sectors for the TPC.
It is required that no energy is found within a wedge of 10$^\circ$ 
opposite to the direction of the lepton transverse momentum. 
To reduce the background from $e^+e^- \rightarrow\mathrm{Z}ee$ with Z decaying to neutrinos,
events are rejected if an electron candidate track is identified and
its energy, including the neutral energy in a 10$^{\circ}$ cone around
the track, is less than 20\,GeV.\par

The selection efficiencies for the three final states are 75\%
(\evev), 77\% (\evmuv) and 43\% (\evtauv). The main background source
is $\mathrm{Z}ee$ where the Z decays to $\nu_\mu \bar{\nu}_\mu$ or
$\nu_\tau \bar{\nu}_\tau$ (the $\nu_\e \bar{\nu}_\e$ case is a
four-fermion final state which is $\eew$ like and is part of the signal).\par

\begin{figure}[H]
   \begin{center}
   \begin{picture}(350,500)
        \put(-415,-80){\mbox{\epsfig{file=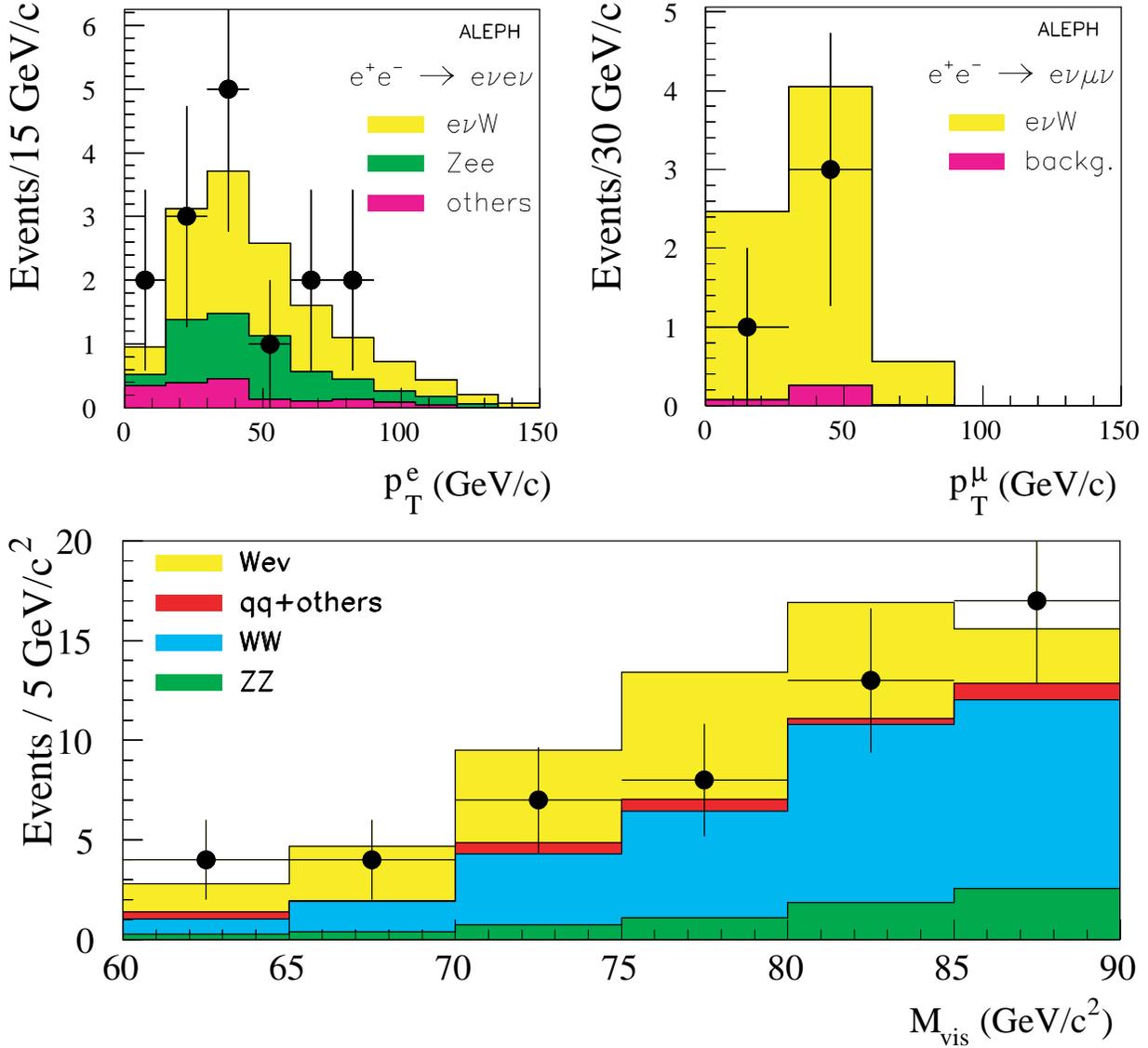,scale=0.95}}}
\put(140,447){\bf{\Huge ALEPH}}
    \end{picture}
   \end{center}
   \caption{The distribution of the lepton transverse momentum,
        $p_{\mathrm{T}}^l$, for single-W events passing the final selection
   cuts for the leptonic electron (upper left) and muon (upper right)
        W decay. The lower plot shows the visible mass distribution
        from single-W events passing the final selection 
   cuts for the hadronic W decay.
     The data are represented by the closed circles.
     The histograms correspond to the Standard Model prediction.}
   \label{fig:Mvis_enw}  
\end{figure}

In the data, 23 events are observed in agreement with the expectation
from the Standard Model of 26.5 events (17.7 signal events).
The composition is 15 events with an electron (8.4 signal and 6.2
background expected), 4 events
with a muon (6.6 signal and 0.4 background expected) and 4 events with
a tau (2.7 signal and 2.2 background expected).
The distributions of the lepton transverse momentum,
$p_{\mathrm{T}}^l$, for single-W events passing the final selection
cuts for the leptonic electron and muon W decay, are shown
in Figure~\ref{fig:Mvis_enw}.

\subsubsection{Hadronic selection}
For the hadronic W decay, the event topology is characterised by two
acoplanar jets with an invariant mass around that of the W boson. The
selection is the same as in the analysis in Ref.~\cite{ref-singlew-183}. In addition to the cut on missing
momentum direction, at least seven good tracks
are required. Similarly to the leptonic selection, tagged two-photon
events and two-fermion events with initial state radiation are
rejected by demanding that the energy $\E_{12}$ be less than
0.025$\sqrt{s}$. The visible mass is required to exceed $60\,\gevm$
and to be less than $90\,\gevm$ to reject untagged two-photon events
at the low end of the mass spectrum and ZZ events at the high end.\par

Events for which the energy in a wedge of 30$^\circ$ centred on the 
transverse missing momentum direction is greater than 0.1$\sqrt{s}$
are rejected. The acollinearity angle between the two hemisphere
(defined by the event thrust axis) momentum directions is required to
be less than 165$^\circ$.\par

The semileptonic final state (\lvqq) of W-pair production is 
efficiently rejected by requiring that no identified electron or muon 
with an energy of more than 0.05$\sqrt{s}$ be reconstructed. The tau jet
reconstruction algorithm of Ref.~\cite{ref-singlew-183} is used in order to further reject
semileptonic decays of W-pairs which contain a tau lepton. After all
cuts, the semileptonic W-pair production, primarily events with one
tau lepton, remains as the dominant background.\par

The efficiency for the hadronic W channel is about 43\%. In the data,
53 events are observed, in agreement with the Standard Model
expectation of 63.1 events (23.5 signal events). The visible mass
distribution of the selected events is displayed in
Figure~\ref{fig:Mvis_enw}.
\subsection{Results}
Limits on \dkg\ and \lgam\ are derived from the total rate of
single-W events, which is sensitive to the \wwgamma\ couplings.
The upper limit on the single-W signal cross section has been
calculated while varying only one coupling at a time, and the 95\%
C.L. limits on \dkg\ and \lgam\ for the 189~GeV data are 
\begin{displaymath}
   \begin{array}{rcccrlcr}
     -2.09 & < & \dkg & < & 0.20 & \   (\lgam  & = & 0) \\
     -0.77 & < & \lgam  & < & 0.79 & \ (\dkg & = & 0),
   \end{array}
\end{displaymath}
including the systematic uncertainties. The different
contributions to the systematic errors are summarised in
Table~\ref{tab-singlew_sys}. The total systematic error amounts to
7\%\ for the leptonic and 12\%\ for the hadronic channel on the
predicted numbers of signal events.
\begin{table}[t]
\caption{\footnotesize Summary of the relative systematic uncertainties in the expected
numbers of selected signal events of the leptonic and the hadronic channels from the
single-W analysis at 189~\gev.}
\begin{center}
\begin{tabular}{|l|c|c|}
\hline
Source & $\Delta N^{exp}_{lep}/N^{exp}_{lep}$ & $\Delta N^{exp}_{had}/N^{exp}_{had}$ \\ \hline \hline
Luminosity              &   $ \pm 0.01 $  &  $\pm 0.03  $ \\
Calorimeter calibration   &   $ - $  &  $^{+0.11}_{-0.08}  $ \\
$\E_{12}$ inefficiency                 &   $ \pm 0.01 $  &  $ -  $ \\
Signal and background cross-section             &   $ \pm 0.06  $ &
$\pm 0.05  $ \\
Fragmentation & $ - $ & $\pm 0.05  $ \\ \hline
Total       &  $\pm 0.07 $ & $ ^{+0.13}_{-0.11} $\\ \hline
\end{tabular}
\label{tab-singlew_sys}
\end{center}
\end{table}
The overall systematic errors are small compared
to the statistical precision, which amounts to 33\%\ for the leptonic
channel and 55\%\ for the hadronic channel.

\begin{figure}[H]
   \begin{center}
   \begin{picture}(350,500)
        \put(-70,-70){\mbox{\epsfig{file=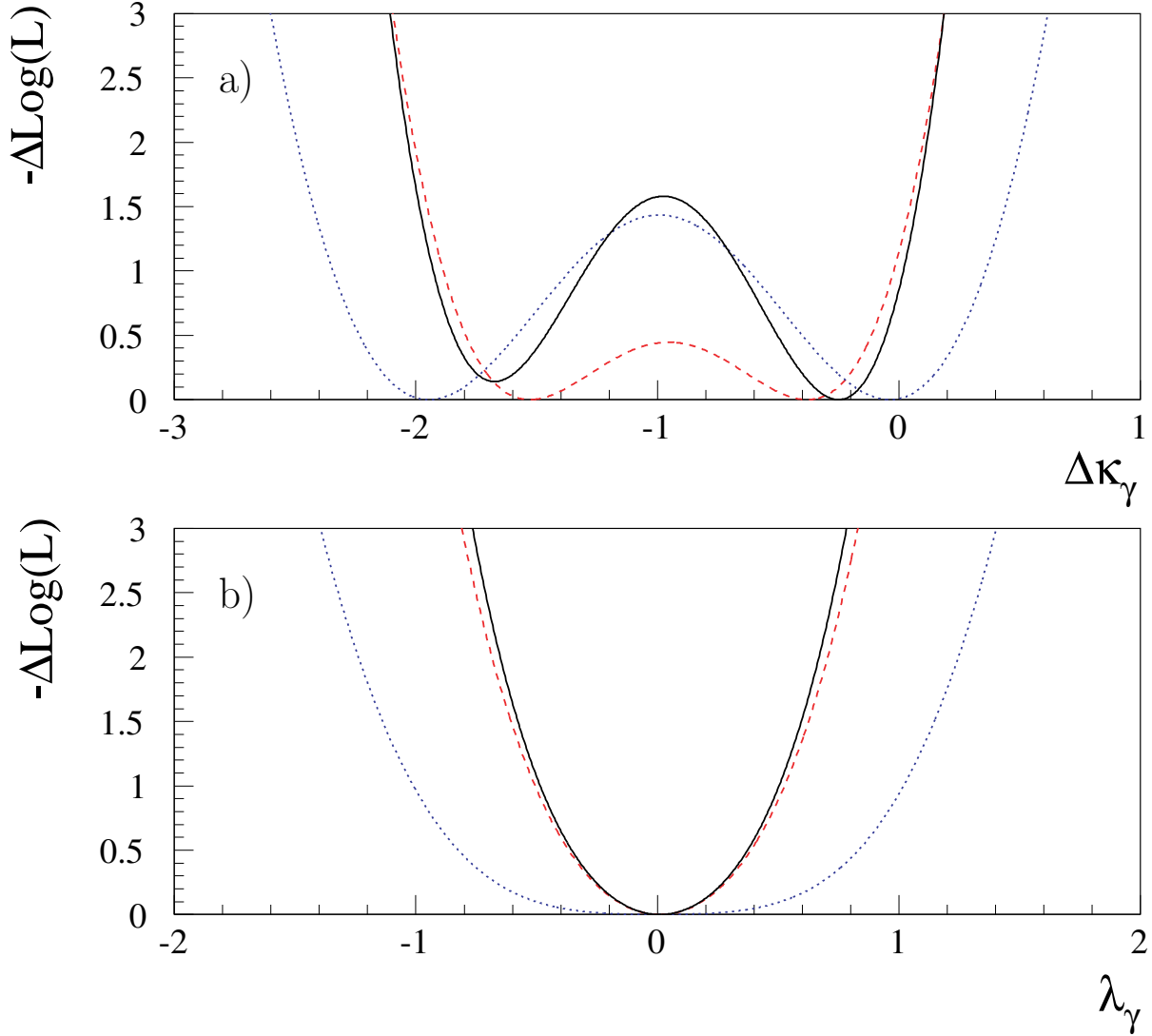,scale=0.9}}}
\put(140,447){\bf{\Huge ALEPH}}
\put(25,360){\Large a)}
\put(25,155){\Large b)}
   \end{picture}
   \end{center}
   \caption{The negative log-likelihood curves from the single-W
analysis as functions of 
            a) $\dkg$ and b) $\lgam$ for the
            189~\gev\ data (dashed line),
            161 - 183~\gev\ data~\cite{ref-singlew-183} (dotted line) 
            and the combined results (solid line). The curve for each
   coupling is obtained while fixing the other coupling to its
   Standard Model value. 
            Systematic errors are not included in these curves.}
   \label{fig:enw_lnl}  
\end{figure}

Combining with the previous measurement for centre-of-mass energies
between 161 and 183~GeV~\cite{ref-singlew-183}, the 95\% C.L. limits
on \dkg\ and \lgam\ from single-W are
\begin{displaymath}
   \begin{array}{rcccrlcr}
     -2.12 & < & \dkg & < & 0.23 & \ (\lgam  & = & 0), \\
     -0.76 & < & \lgam  & < & 0.78 & \ (\dkg & = & 0). 
   \end{array}
\end{displaymath}
The corresponding $\log L$ curves are shown in
Figure~\ref{fig:enw_lnl} for $\dkg (\lgam=0)$ and $\lgam (\dkg=0)$ for
the 189~\gev\ data, 161 - 183\,GeV data and the
combined results.
\section{W-Pair production analysis}
\label{sec-ww}
The large number of W-pair events produced yield the dominant
sensitivity to the TGCs. The
process is sensitive to both the
\wwgamma\ and the \wwz\ couplings via the s-channel W-pair
production diagrams and the sensitivity to the coupling \lgam\ is
higher than that of the single-W and the single-$\gamma$ processes.
\subsection{Event selection and kinematic reconstruction}
\label{subsec-ww-selection}
In this section the event selections for the three distinct \ww\ event
topologies, \lvqq, \qqqq, and \lvlv, are described. Selected events
are exclusively classified in the following order of priority: \muvqq,
\evqq, \qqqq, \tauvqq, and \lvlv.
The expected numbers of events after all cuts used in the TGC
results for signal and background processes at both 
centre-of-mass energies are summarised in Table~\ref{tab-selection-ww} for 
each channel, along with the corresponding selection efficiencies and 
purities.\par
\boldmath
\subsubsection{W$^+$W$^-\rightarrow e\nu q \bar{q}$ and
  W$^+$W$^-\rightarrow \mu\nu q \bar{q}$ events}
\unboldmath
\label{sec-lvqq}
The event selection procedure for semileptonic \ww\ events is similar
to that used for the $\W$ mass measurement at the
corresponding energy~\cite{ref-wmass-183,ref-wmass-189}. At 183~\gev,
events are reconstructed such that they contain a high energy lepton
candidate and two jets~\cite{ref-wmass-183}. The charged particle with
the highest momentum component anti-parallel to the missing momentum
is chosen as lepton candidate. At 189~\gev\ the selection criteria for
the lepton track are slightly changed, using the lepton track
isolation~\cite{ref-wmass-189}. The {\tt
  DURHAM-PE}~\cite{ref-wmass-183} clustering
algorithm is applied to all energy flow objects not used to
construct the lepton four-momentum, and these are forced into two
jets. After this preselection, the probability for the event being signal is 
determined using the momentum of the lepton, the total missing transverse 
momentum and the lepton isolation from the closest jet.\par

At this stage events passing a cut on the probability are
subjected to a kinematic fit in order to improve the resolution on the
reconstructed four-momenta of the W decay products. The kinematic fit
and additional reconstruction cuts are described in the context of the
specific TGC analysis.

\begin{figure}[H]
\begin{center}
\begin{picture}(350,530)
\put(-75,-10){\mbox{\epsfig{file=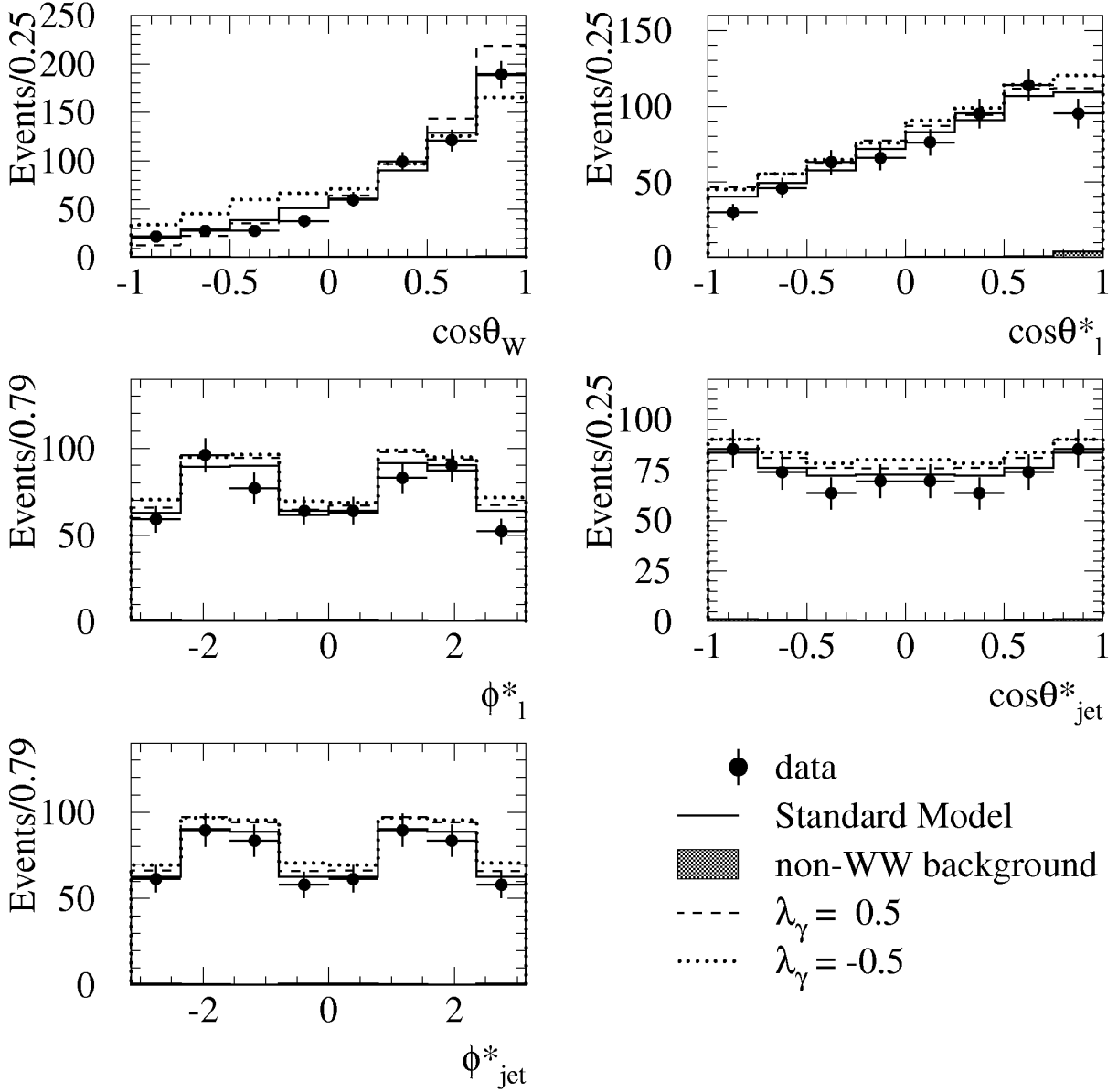,bbllx=28,bburx=584,bblly=118,bbury=674,scale=0.9}}}
\put(140,477){\bf{\Huge ALEPH}}
\end{picture}
\end{center}
\caption[]
{The distributions of the kinematic quantities $\cos\theta_W$,
$\cos\theta_l^*$, $\phi_l^*$, $\cos\theta_{\mathrm{jet}}^*$ and
$\phi_{\mathrm{jet}}^*$  from the combined sample of the \evqq\ 
and \muvqq\ channels at 189~\gev. The measured variables are
the angle $\theta_W$ between the  $\mathrm{W^-}$ and initial $e^-$ in the 
$\mathrm{W^+W^-}$ rest frame, the polar and azimuthal angles of the lepton, 
$\theta^\ast_{\rm l}$ and $\phi^\ast_{\rm l}$, in the rest frame of its parent W, and the polar 
and azimuthal angles of a quark jet, ${\theta}^{\ast}_{\rm jet}$ and ${\phi}{^\ast}_{\rm jet}$, 
in the rest frame of its parent W.  As no quark flavour tagging is performed
each of the two ambiguous solutions enters with a weight
of 0.5.  The data are represented by solid dots, while the
solid and dashed histograms show distributions for Standard Model and
non-standard values of $\lgam=\pm 0.5$.}
\label{fig-lvqqchannel}
\end{figure}

A W-pair event can be characterised by the five measured
angles, $\theta_W$, the $\mathrm{W^-}$ production angle between the
$\mathrm{W^-}$ and initial $e^-$ in the $\mathrm{W^+W^-}$ rest frame,
the polar and azimuthal
angles of the lepton, $\theta^\ast_{\rm l}$ and $\phi^\ast_{\rm l}$,
in the rest frame of its parent W and the polar and azimuthal angles
of a quark jet, ${\theta}^{\ast}_{\rm jet}$ and ${\phi}{^\ast}_{\rm
jet}$, in the rest frame of its parent W. The distributions of the five
angles $\cos\theta_W$, $\cos\theta_l^*$, $\phi_l^*$,
$\cos\theta_{\mathrm{jet}}^*$ and $\phi_{\mathrm{jet}}^*$, for \evqq\
and \muvqq\ events at 189~\gev\ after selection and reconstruction is
represented in Figure~\ref{fig-lvqqchannel}.
%
%
\begin{table}[t]
\caption{\label{tab-selection-ww} {\footnotesize The numbers of events
    after all
cuts applied in the final $\mathrm{W^+W^-}$ TGC results for data and Monte Carlo simulation
in all channels at centre-of-mass energies of 183 and 189~\gev. The
number of Monte Carlo
events is normalised to the respective integrated luminosity of the 
data. The quoted efficiencies $\epsilon$ and purities $p$ are determined from CC03
events with $m_{\mathrm W}=80.35$ \gev$/\mathrm{c^2}$. For a given \ww\
channel, contributions from other channels are considered as background.}}
\begin{center}
\begin{tabular}{|l|c|c|c|c|c||c|c|c|c|c|} \hline

&\multicolumn{5}{|c||}{$\sqrt{s} = 183$~\gev} &\multicolumn{5}{|c|}{$\sqrt{s} = 189$~\gev}\\ \hline

                     & \evqq & \muvqq & \tauvqq & \qqqq & \lvlv & \evqq & \muvqq & \tauvqq & \qqqq & \lvlv \\ \hline \hline


$\mathrm{N_{exp}}$   &105.6&107.1&90.5 &296.8 &29.7 &360.6 &369.9 &230.2& 1202.9 & 101.3\\

$\mathrm{N_{data}}$  &117  &95   &88   & 314  &29   &361   &370   &224  & 1130 & 102\\ \hline \hline

$\epsilon$ $(\% )$   &76.2 &79.5 &50.7 &67.0  &61.9 &74.0  &78.1  &44.8 & 78.6   & 61.9 \\

$p$ $(\% )$          &96.3 &97.8 &73.5 & 90.0  &89.8 &96.7  &98.1  &80.6 & 86.2   & 90.3 \\ \hline

\end{tabular}
\end{center}
\end{table}
\boldmath
\subsubsection{\ww$\rightarrow$\tauvqq\ events}
\unboldmath
The event selection is based on two complementary algorithms,
developed for the cross section measurement at
161~\cite{ref-wxsec-161} and 172~\gev~\cite{ref-wxsec-172}, but
modified to account for the change in event kinematics with
centre-of-mass energy. In summary, events passing a set of
preselection cuts are selected as semileptonic $\tau$ candidate events
if they fulfil either a global or topological selection. The {\tt
DURHAM-PE} clustering algorithm is then applied to all energy flow
objects that are not used to construct the tau four-momentum, and
these are forced into two jets. More detailed descriptions of the
selection and tau reconstruction can be found in the publications on
the W mass determination at the corresponding
energies~\cite{ref-wmass-183,ref-wmass-189}.\par

To improve the resolution of the angular observables a 3-constraint 
kinematic fit is applied, requiring four-momentum conservation and 
reference mass constraints. In the kinematic fit the 
direction of the $\tau$ is approximated by its visible decay 
products and the extra energy loss is compensated by  correction 
coefficients obtained from Monte Carlo simulated \ww$\rightarrow$\tauvqq\ 
events. For single prong $\tau$ decays the charge of the $\tau$ is directly
accessible, but in the case of three-prong $\tau$ decays ambiguities
arise due to mis-assigned particles from the jets to the $\tau$. For
three-prong $\tau$ decays the charge of the $\tau$ is therefore
determined from the sign of the pseudorapidity-weighted average jet charge
of the $\tau$ decay products (Section~\ref{sec-qqqq}), where the
pseudo-rapidity is defined with respect to jet-axis. The charge
mis-assignment
in \tauvqq\ events is 5\%\ for one-prong and 41\%\ for three-prong
$\tau$ decays.

\begin{figure}[H]
\begin{center}
\begin{picture}(350,520)
\put(-60,225){\mbox{\epsfig{file=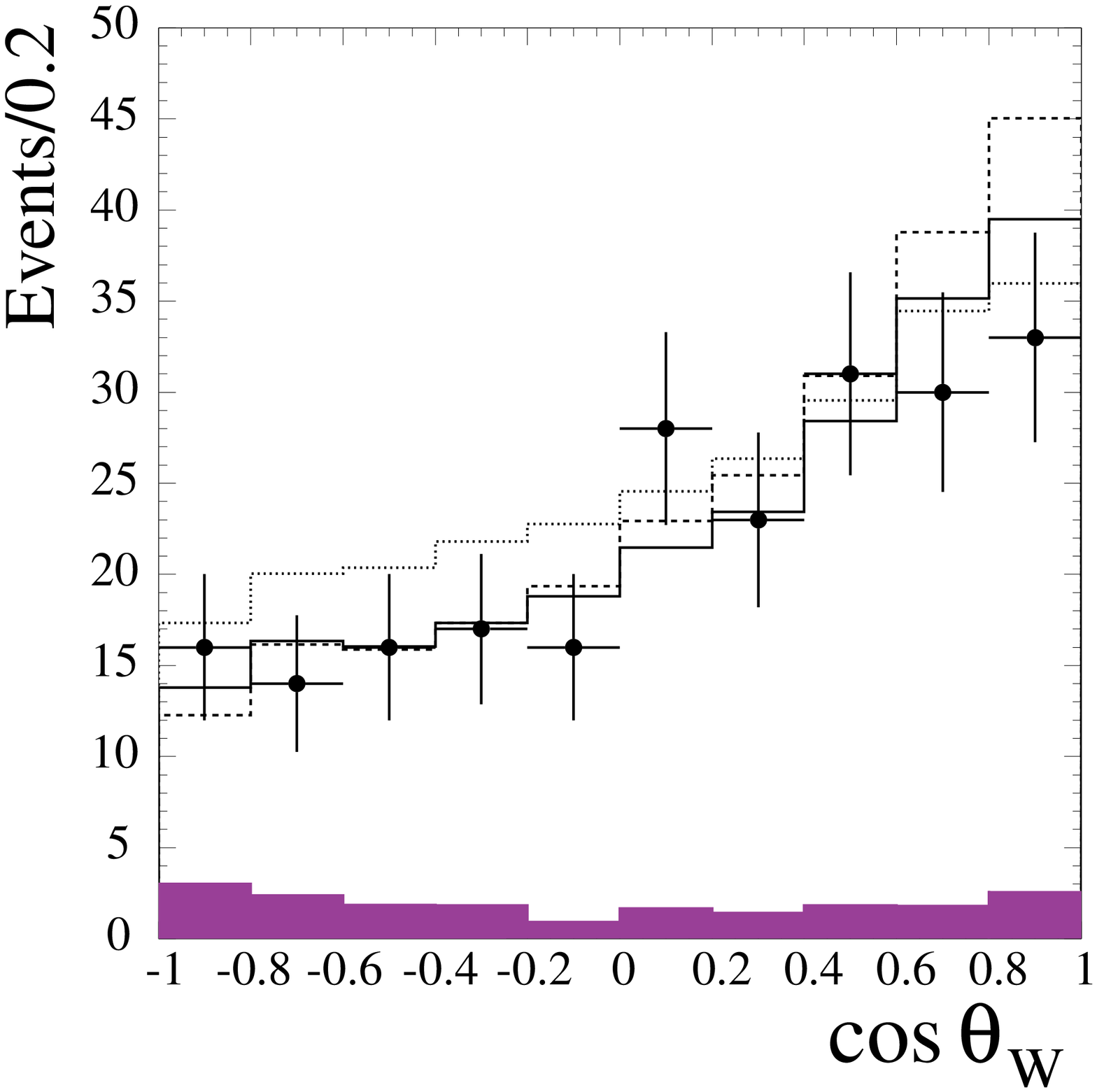,scale=.4}}}
\put(175,220){\mbox{\epsfig{file=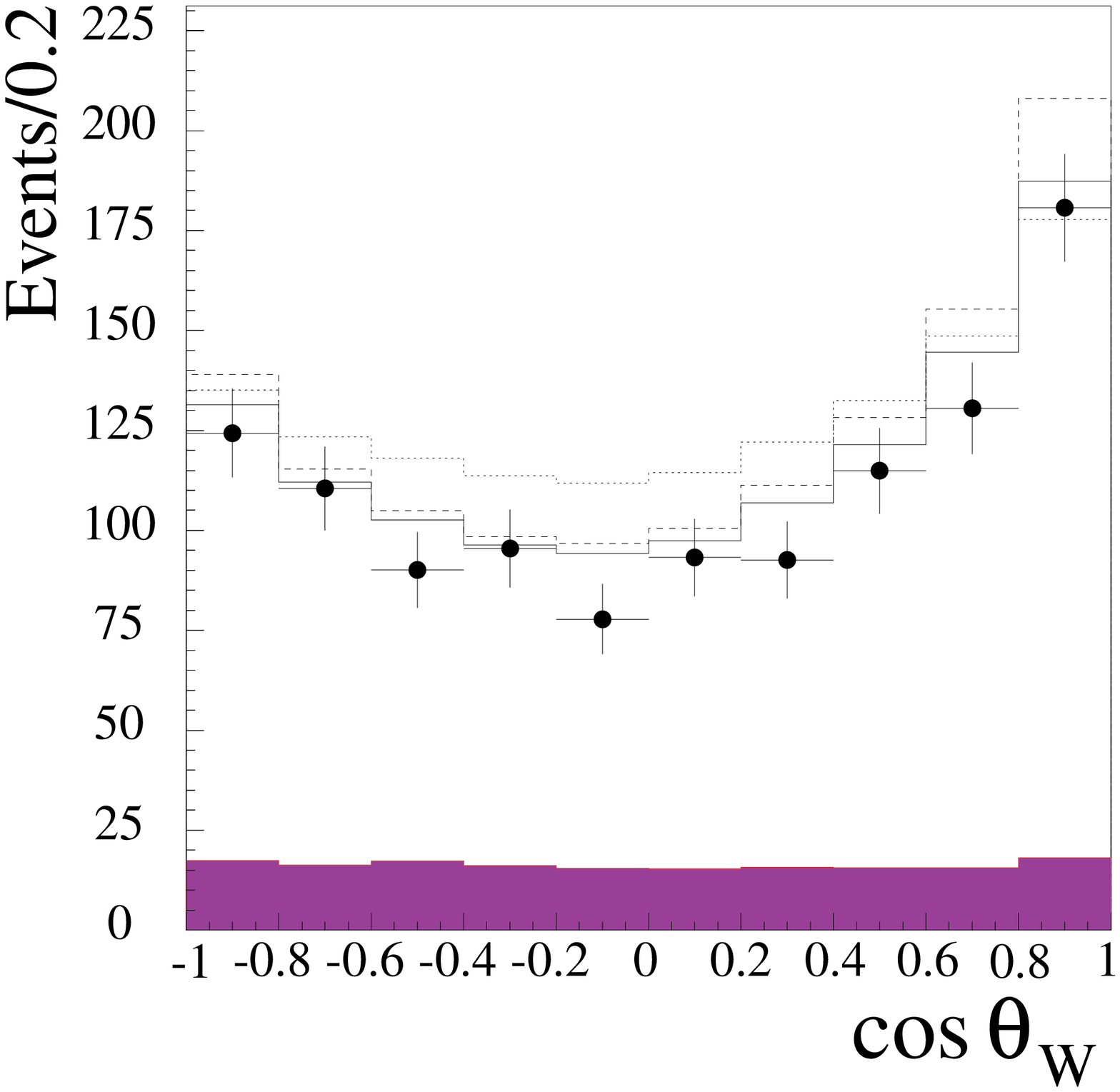,scale=.405}}}
\put(-60,0){\mbox{\epsfig{file=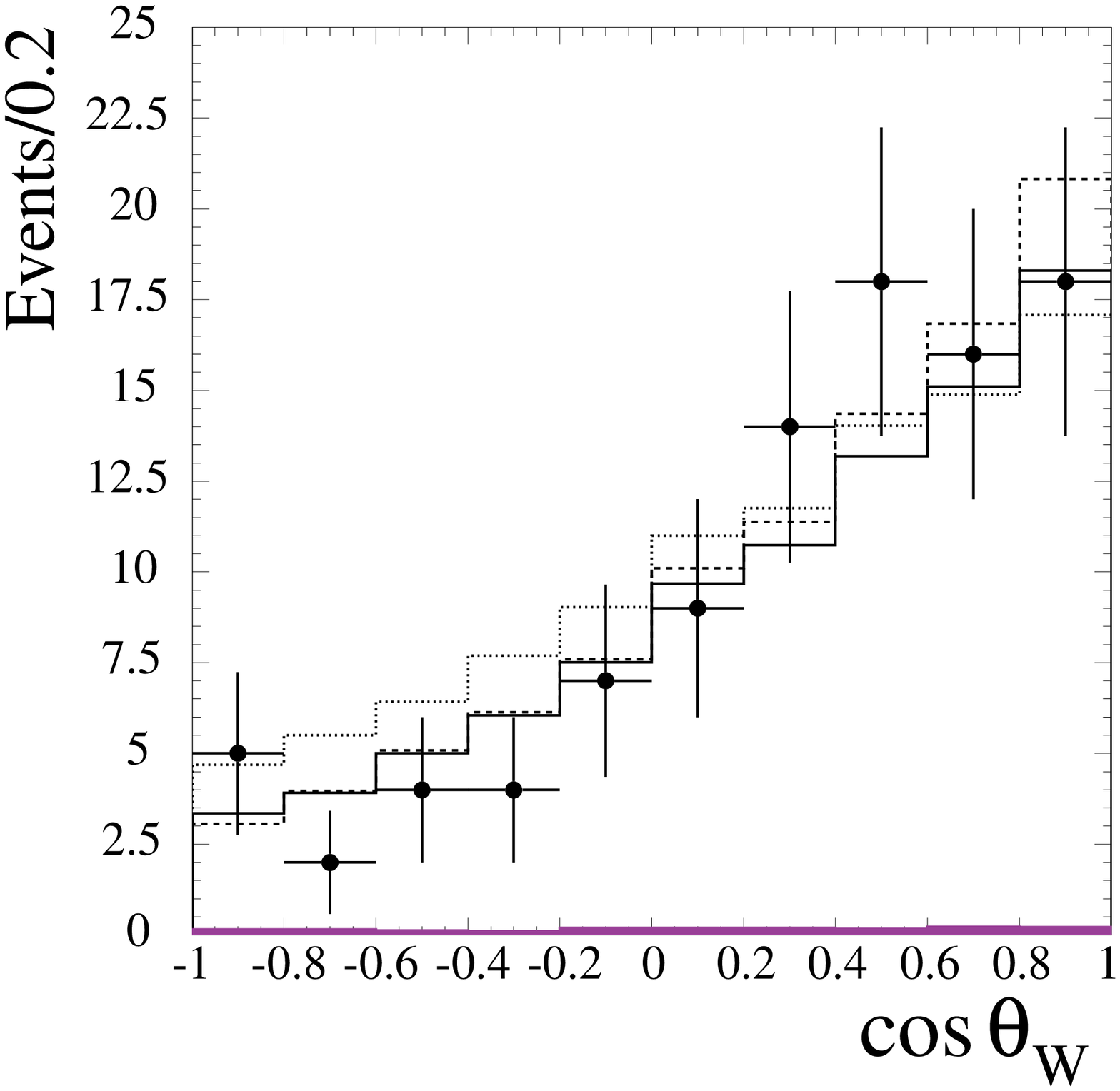,scale=.4}}}
\put(50,-160){\mbox{\epsfig{file=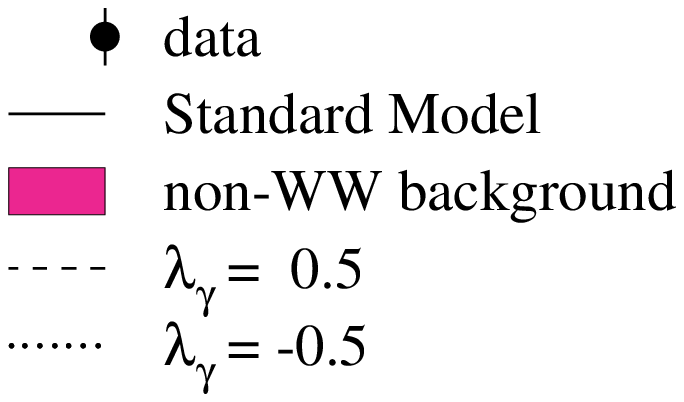,scale=1.0}}}
\put(-5,415){\Large a)}
\put(229,415){\Large b)}
\put(-5,190){\Large c)}
\put(140,467){\bf{\Huge ALEPH}}
\end{picture}
\end{center}
\caption[]
{Distributions of the cosine of the W$^-$ production angle,
  $\cos\theta_W$, at 189 GeV for a)~\tauvqq, b)~\qqqq\ and c)~\lvlv\ events. The
  data are represented by solid dots, while the solid and dashed
  histograms show distributions for Standard Model and non-standard 
  values of the TGCs. The shaded area represents the non-WW
 background.  For \qqqq\ events, each event
  enters with two solutions for $\cos\theta_W$ in the distribution with
  the weights $P_+$ and $1-P_+$, where $P_+$ is the
  probability for a di-jet pair to be a W$^+$. For \lvlv\ events, each event
  enters with two solutions for $\cos\theta_W$ in the distribution with
  a weight of 0.5.}
\label{fig-otherchannels}
\end{figure}

The distribution of the cosine of the W$^-$ production angle from \tauvqq\
events can be seen in Figure~\ref{fig-otherchannels}.
\boldmath
\subsubsection{\ww$\rightarrow$\qqqq\ events}
\unboldmath
\label{sec-qqqq}
To extract the hadronic \ww\ signal with high purity and efficiency,
the selection is based on a neural
network~\cite{ref-wxsec-183}. Events passing a preselection designed 
to remove the q$\overline{\mathrm q}(\gamma)$ background, are assigned
a neural network output, based on global event properties, heavy quark
flavour tagging, jet properties and WW kinematics. A detailed
description of the selection algorithm at 183 and 189~\gev\ is given
in~\cite{ref-wxsec-183}.\par

For the hadronic \ww\ events the reconstruction of the relevant
information is more complicated since the W$^-$ direction is not known
and the information on the particle flavours in
either W system is not discriminant. In this case the four jets can
be paired in three
different ways. To select the best pairing, a 6-constraint kinematic
fit is applied to all three possible pairings. The kinematic fit
requires four-momentum conservation and reference mass constraints.
The four-momenta  
obtained in the kinematic fit for the pairing with the lowest $\chi^2$
value are then used in the final determination of the TGCs, while the
other combinations  are discarded. The efficiency of this algorithm to
find the correct combination was found to be 78\%\ at 183 and 75\%\ at
189~\gev. 

\begin{figure}[H]
\begin{center}
\begin{picture}(350,535)
\put(-83,-55){\mbox{\epsfig{file=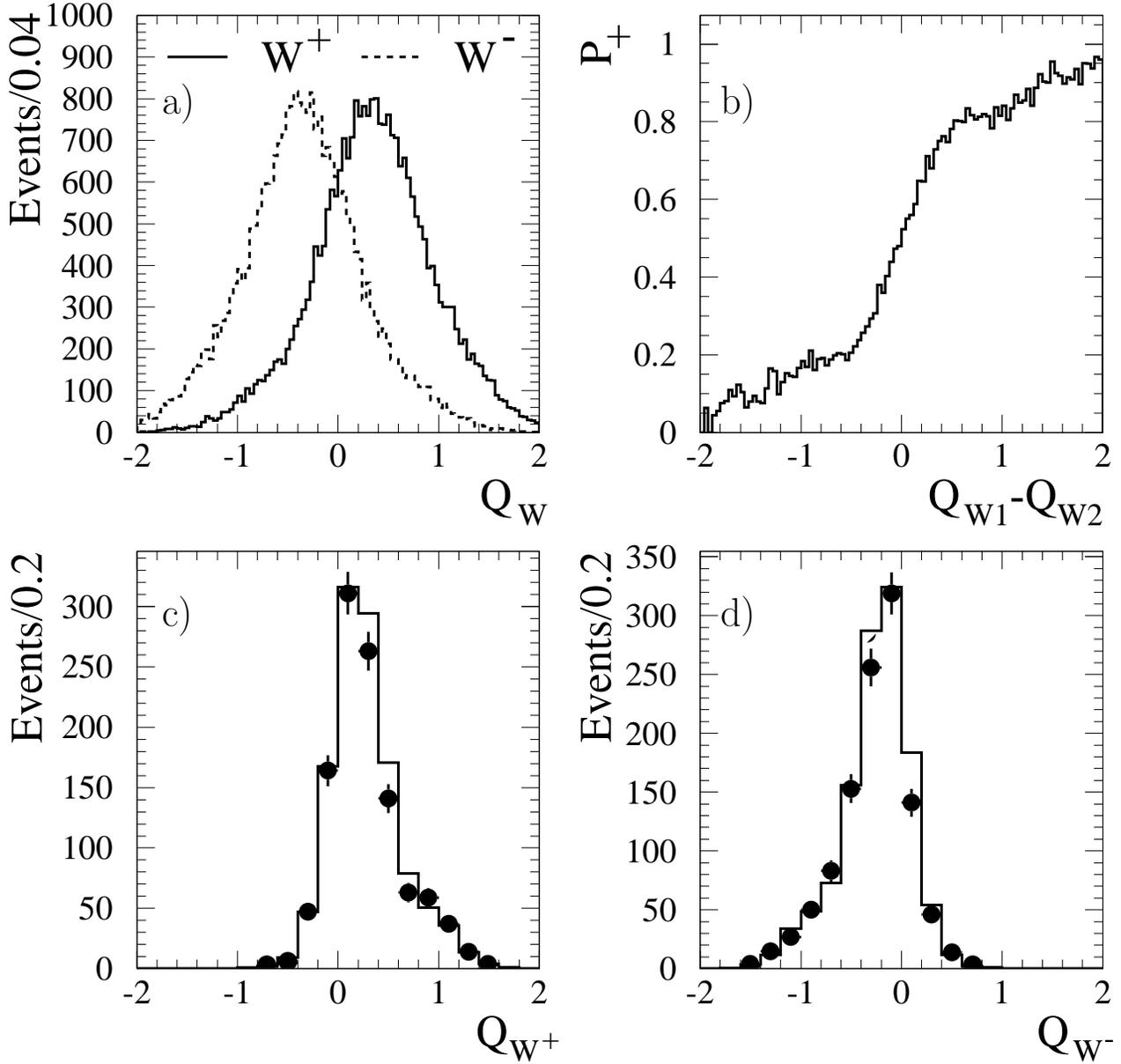,scale=0.9}}}
\put(140,482){\bf{\Huge ALEPH}}
\put(0,410){\Large a)}
\put(240,410){\Large b)}
\put(0,190){\Large c)}
\put(240,190){\Large d)}
\end{picture}
\end{center}
\caption[]
{W-charge tagging distributions from \qqqq\ W-pair events at
189~\gev. a) The distribution of the jet pair charge for W$^+$
(solid histogram) and W$^-$ (dashed
  histogram) decays for Monte Carlo events. b) The
  probability $P_+$ as function of the charge difference between the
  two W's. c,d) Experimental distributions of $Q_{{\mathrm W}^+}$ and
$Q_{{\mathrm W}^-}$  from semileptonic events. The data are
represented by the dots and the Monte Carlo simulation by the histograms. The
number of Monte Carlo events is normalised to the integrated luminosity of data.}
\label{fig-charge}
\end{figure}

To assign a jet pair to the W$^+$ or W$^-$ a jet charge algorithm is
used. The jet charge, $Q_{\rm jet}$, is obtained from
the pseudorapidity-weighted average charge of jet particles.
The jet
pair charge is defined by the sum of the two jets assigned to a W,
$Q_{\mathrm W}=Q_{\rm jet1}+Q_{\rm jet2}$. A jet pair is then assigned
to the W$^+$ based on the charge difference between the two jet pairs,
$\Delta Q$, with a probability $P_{+}$. The probability $P_{+}$ is
given by
\begin{equation}
 P_{+}(\Delta Q) = \frac{{\cal N}_+(\Delta Q)}{ {\cal N}_+(\Delta Q)
 + {\cal N}_+(-\Delta Q) },
\label{eq-qdijet}
\end{equation}
where ${\cal N}_+$ is the probability density function for the
charge difference 
between the two W systems for true W$^+$ jet pairs obtained from Monte
Carlo event samples~\cite{ref-wwtgc-172}. Figure~\ref{fig-charge}
shows the distribution of the di-jet charge of the two W systems for
true W$^+$ and W$^-$ decays, obtained from Monte Carlo
generator information. The distribution of the di-jet charge from
semileptonic events is also shown for both data and Monte Carlo. The
data are well reproduced by the Monte Carlo simulation. The charge assignment
efficiency for correctly paired hadronic \ww\ events amounts to
approximately 76\%\ for $P_+ > 0.5$.

The distribution of the cosine of the W$^-$ production angle from
\qqqq\ events can be
found in Figure~\ref{fig-otherchannels}.
\boldmath
\subsubsection{\ww$\rightarrow$\lvlv\ events}
\unboldmath
The selection of \ww$\rightarrow$\lvlv\ events (where $\ell$ denotes
an electron or muon) is mainly based on variables used
in Ref.~\cite{ref-wxsec-161}, namely missing transverse momentum, missing
mass and kinematic properties of the lepton candidate. The information
of these and other variables is combined in a neural network. A detailed
description of all the variables used in the neural network can be found
in the appendix~\ref{sec-nnvarlep}.\par

For purely leptonic \ww\ events the momenta of the two neutrinos are
unknown. However, in the absence of ISR and neglecting the W width,
the constraint that the two $l\nu$ systems should have the W mass
($M_1 =M_2 =80.35$~\gev$/c^2$) in
combination with the usual four-momentum conservation allows a
reconstruction of the neutrino momenta. The quadratic nature of the
mass constraint results in a two-fold ambiguity, corresponding to
flipping both neutrinos with respect to the plane defined by the
charged leptons. As detector resolution, ISR and the finite W width are
not included in this reconstruction hypothesis, 28.0\%\ of the events
have no physical solution and a zero-constrained kinematic fit is
employed. The fit determines a set of values for the reference masses,
$M_1$ and $M_2$, in the mass-constraints for the two $l\nu$ systems,
for which a physical solution exists. By this method 92\%\
of the events without a solution are recovered, resulting in a 97.7\%\
reconstruction efficiency for signal events. The majority of the
events which fail to have a solution are purely leptonic \ww\ events
with at least one leptonically decaying $\tau$, which is the dominant
background. In addition to the
selection by a cut on the neural network output, purely leptonic
events are only accepted if a physical solution is found inside the
mass window of $55-105$ \gev$/c^2$.\par 

The distribution of the cosine of the W$^-$ production angle for
\lvlv\ events at
189~\gev\ after selection and reconstruction is represented in  
Figure~\ref{fig-otherchannels}.
\subsection{Determination of the TGCs}
Three different methods, described in the following, are used to
extract the couplings in the different W-pair final states.
\label{subsec-determination}
\subsubsection{The optimal observable methods}
The general idea of optimal observables (OO)~\cite{ref-oo} is to
project the sensitive kinematic information for a given coupling $g_i$
onto the one-dimensional distribution of a suitably defined variable
${\cal O}_i^{(1)}$. The coupling $g_i$ can then be extracted from a fit to
this distribution or, equivalently, from the measurement of the mean
value $\langle{\cal O}_i^{(1)} \rangle$ of the optimal observable.\par

Since the amplitudes are linear in the TGCs the differential cross
section can be expanded in these couplings $g_i$ containing no terms
beyond the second order 
\begin{equation}
  \dsig =  \sz (1 + \sum_i \oi^{(1)} \cdot g_i + \sum_{ij} {\cal O}_{ij}^{(2)} \cdot g_i g_j) ,
\label{diffxsec}
\end{equation}
where $g_i$ denotes any type of couplings and $\Omega$
denotes phase space variables taking into account reconstruction
ambiguities for the individual \ww\ channels. The zero-order
term, $\sz$, represents the Standard Model contribution.
Using the first order term, a given set of couplings, $\overline{g}$,
can be determined by minimising 
\begin{equation}
\chi^2 (\overline{g}) = \sum_{ij}(\langle {\cal O}_i^{(1)} \rangle - E[{\cal
  O}_i^{(1)}])V({\cal O})_{ij}^{-1}(\langle {\cal O}_j^{(1)} \rangle - E[{\cal
  O}_j^{(1)}]),
\label{chi2oo}
\end{equation}
where ${\langle \cal O}_i^{(1)} \rangle $ and $V({\cal O})$ are the measured
mean values and their covariance matrix. The expected mean values,
$E[{\cal O}_i^{(1)}]$, are obtained by reweighting of fully simulated \ww\
events.

In order to ensure maximal sensitivity, $g_i$ can be determined by two
different approaches:
\begin{itemize}
\item An iterative procedure (denoted {\bf OO$_1$} in the further
text), where the cross section in Eq.~\ref{diffxsec} is expanded about
a given coupling value and consequently the definition of the
observable ${\cal O}_i^{(1)}$ is re-optimised.
\item Adding the information contained in the second order term of the
expansion in Eq.~\ref{diffxsec} (called {\bf OO$_2$} in the further
text). This is achieved by including the second optimal observable and
adding additional terms of
the same structure as the ones in Eq.~\ref{chi2oo} to the
$\chi^2$, including new terms describing the correlation between
${\cal O}_i^{(1)}$ and ${\cal O}_{ij}^{(2)}$. The second order
observable increases the sensitivity when the information contained
in the first order observable decreases~\cite{ref-oo2}. The
covariance matrix for the mean values are obtained by reweighting
fully simulated \ww\ events.
\end{itemize}
In both cases the information from the measured cross section is
included by adding a Poisson term to the likelihood function.\par

By construction, these methods are bias-free and take into account any
experimental effect, provided that the Monte Carlo simulation
describes the data correctly. For a given channel, contributions from
any other channel are considered as coupling dependent backgrounds. In
addition the efficiencies and purities of each selection are
parameterised as function of the couplings.\par

For semileptonic events, both OO analyses, OO$_1$ and
OO$_2$, apply a two-constraint kinematic fit using four-momentum
conservation, equal W-mass hypothesis and including a massless
neutrino. The corresponding $\chi^2$-probability of the fit is
required to be larger than 0.005 for an event to be selected for the
TGC extraction. This cut improves the purity of the sample and
discards poorly reconstructed events. In addition, an OO-window
cut is applied, optimised to improve the sensitivity to the TGCs by
reducing the contamination of background events with OO values
incompatible with W-pair production.\par

For hadronically decaying W's, there remains a twofold ambiguity since
the quark flavours are undetermined. Hence, for semileptonic \evqq,
\muvqq\ and \tauvqq\ events the
contributions are averaged over the quark and anti-quark
directions. For hadronic events, considering the W charge as undetermined,
there is an eightfold ambiguity. This is included
in the final extraction of the TGCs, where each contribution is weighted with the
corresponding di-jet charge probability, Eq.~\ref{eq-qdijet}. For
leptonic events, \lvlv, the contributions are averaged over the two
solutions for the neutrino momenta.
\subsubsection{Maximum likelihood-fit}
\label{subsec-maximum}
A maximum likelihood analysis (LL), Eq.~\ref{eq-lldef}, of the channels \evqq\ and
\muvqq\ is used to measure the C- or  P-violating couplings and as a
cross-check for the CP-conserving couplings. The
measured variables are the five angles described in Sect.~\ref{sec-lvqq}. As
in the optimal observable analyses, no quark flavour tagging is
performed and the quark and anti-quark directions are averaged. The
probability density function, $P$, is given by 
\begin{equation}
    P(\bar{\Omega},\bar{g}) = {b(\bar{\Omega})+s(\bar{\Omega},\bar{g}) \over B + S(\bar{g})},
\label{eq:pdf}
\end{equation}
where $\overline{g}$ denotes a set of couplings and the angles $\bar{\Omega}= (\theta_W,\theta^\ast_{\rm
l},\phi^\ast_{\rm l},\theta^\ast_{\rm jet}, \phi^\ast_{\rm jet})$ are
calculated using the charged lepton, neutrino, and quark jet
four-vectors. The
quantity $b(\bar{\Omega})$ is the background distribution as predicted by
Monte Carlo. The $\mathrm{W^+W^-}$ signal distribution,
$s(\bar{\Omega},\bar{g})$, is defined by
\begin{equation}
    s(\bar{\Omega},\bar{g}) = \int d \hat{s}\, d \bar{\Omega}^{\mathrm{true}}\, r(\bar{\Omega},\bar{\Omega}^{\mathrm{true}})\, \epsilon(\hat{s},\bar{\Omega}^{\mathrm{true}})\, F(\hat{s})\, \frac{d\sigma}{d\bar{\Omega}'}(\hat{s},\bar{\Omega}^{\mathrm{true}},\bar{g}),  \label{eq:xmu}
\end{equation}
where $\hat{s}$ is the squared invariant mass of the $W^+W^-$ system,
$r(\bar{\Omega},\bar{\Omega}^{\mathrm{true}})$ is the detector resolution
function, $\epsilon(\hat{s},\bar{\Omega}^{\mathrm{true}})$ is the detection
efficiency, $F(\hat{s})$ is an initial state radiation
function~\cite{ref-fadin}, and
$\frac{d\sigma}{d\bar{\Omega}'}(\hat{s},\bar{\Omega}^{\mathrm{true}},\bar{g})$
is the lowest-order narrow-width differential cross-section for
$\mathrm{W^+W^-}$ production and decay\cite{ref-hagivara}.\par

The normalisation factors $B$ and $S(\bar{g})$ are the integrals of
$b(\bar{\Omega})$ and $s(\bar{\Omega},\bar{g})$,
where $S(\bar{g})$ is evaluated by reweighting $W^+W^-$ Monte Carlo
events in order to include detector resolution and efficiency. The
proper evaluation of the normalisation $S(\bar{g})$ is crucial for the
success of the likelihood method.\par

Some approximations to $s(\bar{\Omega},\bar{g})$ are made when
evaluating the numerator of Eq.~\ref{eq:pdf}.  In particular, most of the
information in the detection efficiency function
$\epsilon(\hat{s},\bar{\Omega})$ arises 
from the charged lepton momentum $p_l$ and polar angle $\theta_l$ in
the laboratory reference frame. This dependence has been
parameterised with a two-dimensional efficiency function.

The event selection for this method is
the same as for the optimal observables. A kinematic fit using the
equal mass hypothesis and four-momentum conservation is
applied, where the corresponding $\chi^2$-probability
is required to be larger than 0.02.
The efficiencies, purities and numbers of events for
data and Monte Carlo simulation are shown in
Table~\ref{tab-selection-maxlik}.
\begin{table}[t]
\caption{ \label{tab-selection-maxlik} {\footnotesize The numbers of events after all
cuts applied in the  maximum likelihood TGC analysis for
data and Monte Carlo simulation at centre-of-mass energies of 183 and
189~\gev. The number of Monte Carlo events is normalised to the
respective integrated luminosity of the data. The quoted efficiencies
$\epsilon$ and purities $p$ are determined from CC03 events with
$m_{\mathrm W}=80.35$ \gev$/\mathrm{c^2}$. Only non-\ww\ events are
considered as background in the calculation of the efficiencies
$\epsilon$ and purities $p$.}}
\begin{center}
\begin{tabular}{|l|c|c||c|c|} \hline
  & \multicolumn{2}{c||}{$\sqrt{s} = 183$~\gev} & \multicolumn{2}{c|}{$\sqrt{s} = 189$~\gev} \\ \hline
    Channel          & \evqq    & \muvqq  & \evqq   & \muvqq  \\ \hline \hline
$\mathrm{N_{exp}}$   & \parbox{2.75cm}{\center 91.8}  & 97.4 & \parbox{2.85cm}{\center 293.0}  & 311.7 \\
$\mathrm{N_{data}}$  & 98    & 86   & 275  & 310    \\ \hline \hline
$\epsilon$ $(\% )$   & 66.4  & 71.0 & 66.1 & 70.9 \\
$p$ $(\% )$          & 98.5  & 99.4 & 98.6 & 99.5 \\ \hline
\end{tabular}
\end{center}
\end{table}
\subsection{Results}
\subsubsection{Comparison of methods}
\label{sec-comp}
The three methods (OO$_1$, OO$_2$ and the maximum likelihood)
discussed above are used to determine the TGCs from WW events.\par

The OO$_1$ method is used to measure the couplings \dgz, \dkg\
and \lgam\ for all five WW final states considered in this analysis,
namely \evqq, \muvqq, \tauvqq, \qqqq\ and \lvlv. The OO$_2$ method is
employed to measure the couplings \dgz, \dkg\ and \lgam\ for the
\evqq, \muvqq\ and \qqqq\ channel. The  maximum likelihood
method is
used in the two semileptonic channels, \evqq\ and \muvqq, to measure
the standard set of couplings \dgz, \dkg\ and \lgam\ and the real and
imaginary parts of the C- or P-violating couplings $\mathrm{g^V_4}$,
$\mathrm{g^V_5}$, $\mathrm{\tilde{\kappa}_{V}}$, and
$\mathrm{\tilde{\lambda}_{V}}$, where V denotes either $\gamma$ or Z.\par

A comparison of the three couplings \dgz, \dkg\ and \lgam, using the
three methods OO$_1$, OO$_2$ and  maximum likelihood, is given in
Table~\ref{tab-res-stat}.
For OO$_1$ and OO$_2$, the combined results at 183 and 189~\gev\ are
extracted by adding up the corresponding $\chi^2$ terms, while for
the  maximum likelihood method results are
extracted by summing up the corresponding $\log L$ functions. The
error intervals for each coupling are defined as the 68\%\ confidence
intervals obtained by integration of the likelihood functions, to
accommodate cases with non-parabolic behaviour of the log-likelihood
function.
\begin{table}
\caption{\label{tab-res-stat}\footnotesize Comparison of the three
  couplings \dgz, \dkg\ and \lgam, using the three methods OO$_1$,
  OO$_2$ and  maximum likelihood. The error intervals for each
  coupling are statistical only.}
\begin{center}
\begin{tabular}{|c|c||c|c|c|} \hline
 & & \multicolumn{3}{|c|}{Method} \\ \cline{3-5}
Channel & Coupling & OO$_1$ & OO$_2$ & LL \\ \hline
 \myrule         & \dgz  & $~~0.07_{\T -0.10}^{\T +0.11}$ & $~~0.10_{\T -0.10}^{\T +0.09}$ & $~~0.03_{\T -0.10}^{\T +0.10}$ \\
 \myrule \evqq   & \dkg  & $-0.09_{\T -0.34}^{\T +0.50}$ & $~~0.45_{\T -0.33}^{\T +0.35}$ & $~~0.11_{\T -0.30}^{\T +0.47}$ \\
 \myrule         & \lgam & $~~0.04_{\T -0.11}^{\T +0.12}$ & $~~0.22_{\T -0.10}^{\T +0.11}$ & $~~0.11_{\T -0.11}^{\T +0.11}$ \\ \hline
 \myrule         & \dgz  & $~~0.05_{\T -0.10}^{\T +0.10}$ & $~~0.00_{\T -0.08}^{\T +0.10}$ & $~~0.06_{\T -0.09}^{\T +0.09}$ \\
 \myrule \muvqq  & \dkg  & $-0.02_{\T -0.34}^{\T +0.52}$ & $~~0.24_{\T -0.35}^{\T +0.59}$ & $~~0.38_{\T -0.38}^{\T +0.49}$ \\
 \myrule         & \lgam & $-0.03_{\T -0.09}^{\T +0.10}$ & $-0.08_{\T -0.08}^{\T +0.09}$ & $-0.08_{\T -0.09}^{\T +0.09}$ \\ \hline
 \myrule         & \dgz  & $~~0.51_{\T -0.29}^{\T +0.19}$ & - & - \\
 \myrule \tauvqq & \dkg  & $-0.71_{\T -0.32}^{\T +0.39}$ & - & - \\
 \myrule         & \lgam & $~~0.00_{\T -0.14}^{\T +0.17}$ & - & - \\ \hline
 \myrule         & \dgz  & $-0.06_{\T -0.09}^{\T +0.10}$ & $-0.03_{\T -0.10}^{\T +0.11}$ & -\\
 \myrule \qqqq   & \dkg  & $-0.11_{\T -0.27}^{\T +0.30}$ & $~~0.21_{\T -0.78}^{\T +0.34}$ & -\\
 \myrule         & \lgam & $-0.15_{\T -0.10}^{\T +0.11}$ & $-0.02_{\T -0.12}^{\T +0.14}$ & -\\ \hline
 \myrule         & \dgz  & $-0.17_{\T -0.20}^{\T +0.30}$ & - & - \\
 \myrule \lvlv   & \dkg  & $-0.35_{\T -0.41}^{\T +0.80}$ & - & - \\
 \myrule         & \lgam & $~~0.05_{\T -0.13}^{\T +0.13}$ & - & - \\ \hline
\end{tabular}
\end{center}
\end{table}\par

The linearity of the three fitting procedures is checked by repeating
the fits using Monte Carlo event samples generated with non-zero
values for the TGCs. In all cases the results are consistent to those
values within
the statistical uncertainty of the Monte Carlo samples.

The consistency of the Monte Carlo simulations with the data is
verified by comparisons of the distributions of the input quantities
to the selections for data and Monte Carlo. In addition, the stability
of the analysis with respect to the event selection is tested by
varying the main selection criteria within reasonable limits. In no
case significant discrepancies are found.

The reliability of the errors from each fitting procedure is
investigated by performing fits to a large number of independent Monte
Carlo samples, each corresponding to the integrated luminosity of the
data. These samples, typically 300, are then reweighted to non-zero
values for the TGCs and passed through the analysis chain. The
expected 68\%\ confidence levels, obtained from the distributions of
the fit values, show good correspondence with the 68\%\ confidence
intervals obtained for data.\par

The consistency of the data results from the three different methods used for
the \evqq\ and \muvqq\ channels, has been checked by performing
fits to a large number of independent Monte Carlo samples.
The results obtained with the three
different methods are compatible and the expected spread between
the methods show a good agreement with the observed differences in the
data.

The expected 68\%\ confidence level intervals, obtained
from the distributions of the fit values for the three methods at 189~\gev\ for
the three couplings \dgz, \dkg\ and \lgam, are listed in
Table~\ref{tab-res-cor}. The expected errors of the OO$_2$ and the
 maximum likelihood method are very similar, whereas the
OO$_1$ method is slightly worse in the case of \dkg. For the final results, the OO$_2$
method is therefore employed in the analysis of the \evqq, \muvqq\ and
\qqqq\ final states and the OO$_1$ analysis is used for the remaining
\tauvqq\ and \lvlv\ final states.

\begin{table}
\caption{\label{tab-res-cor} \footnotesize
The expected error for the three methods, OO$_1$,
OO$_2$ and  maximum likelihood, at
189~\gev\ for the three couplings \dgz, \dkg\ and \lgam. The
 maximum likelihood method has only been applied in the two
semileptonic channels, \evqq\ and \muvqq.
}
\begin{center}
\begin{tabular}{|c|c||c|c|c|} \hline
 & & \multicolumn{3}{|c|}{Expected error} \\ \cline{3-5}
Channel & Coupling & OO$_1$ & OO$_2$ & LL \\ \hline
\myrule        & \dgz  & 0.11 & 0.12 & 0.11 \\
\myrule\evqq   & \dkg  & 0.48 & 0.43 & 0.40 \\
\myrule        & \lgam & 0.13 & 0.12 & 0.13 \\ \hline
\myrule        & \dgz  & 0.11 & 0.11 & 0.10 \\
\myrule\muvqq  & \dkg  & 0.48 & 0.41 & 0.37 \\
\myrule        & \lgam & 0.11 & 0.11 & 0.13 \\ \hline
\end{tabular}
\end{center}
\end{table}

\subsubsection{Systematic uncertainties}
\label{sec-syst}
In the following the different sources of systematic errors for each
decay channel and their determination for the various methods are
briefly described. The different contributions of each source to the
total systematic error for the three couplings \dgz, \dkg\ and \lgam,
as obtained with the OO methods, are given for each channel
in Table~\ref{tab-systematics-189}. The systematic
uncertainties for the combined \evqq\ and \muvqq\ channels
are listed in Table~\ref{tab-systematics-189-cp} for the C-
or P-violating couplings, determined with the  maximum
likelihood method.

The following sources, listed in the
approximate relative importance, have been considered to be
fully correlated between the channels:
\begin{itemize}
\item[-]{\bf Fragmentation:}
  The effect of fragmentation in hadronic W decays is estimated
  by fitting the couplings in samples of events generated with {\tt
    KORALW} where the default JETSET fragmentation is replaced
  by {\tt HERWIG} \cite{ref-herwig} fragmentation. The {\tt HERWIG}
  fragmentation parameters are tuned at the Z using hadronic
  events with flavour tagging \cite{ref-wmass-189}.
\item[-]{\bf W$^+$W$^-$ cross section:}
  The uncertainty due to the theoretical error on the expected \ww\ 
  cross section predicted by {\tt KORALW} is estimated by changing the
  \ww\ cross section by $\pm 2$\%~\cite{ref-lep2report}.
\item[-]{\bf Luminosity:}
  The effect of the error on the integrated luminosity is
  estimated by varying the measured integrated luminosity by $\pm
  0.7$\%\ \cite{ref-wxsec-189}.
\item[-]{\bf LEP energy:}
  The uncertainty on the LEP energy affects the determination of the
  couplings via the kinematic fitting procedure and the cross section
  measurement. The values of the LEP energy are varied in the range
  $ \pm 0.050$~\gev\ \cite{ref-lepenergy}, which
  has a negligible effect on the results.
\item[-]{\bf W mass:}
  The analysis is repeated using Monte Carlo samples generated at
  different values of the W mass to investigate the effects due to the
  uncertainty of $\pm 62$~\mev\ in the W mass measured at hadron
colliders~\cite{ref-wmass-TEV}.
\item[-]{\bf Calorimeter absolute scale:}
  The absolute energy scale of the electromagnetic and hadronic
  calorimeters is determined using hadronic Z
  events. The uncertainties in the absolute scale are found to be $\pm
  0.9$\%\ and $\pm 2$\%\ for the electromagnetic and hadronic
  calorimeter, respectively. The effect of a possible miscalibration
  of the calorimeters is evaluated on Monte Carlo samples by scaling
  the electromagnetic and hadronic part of the measured energy
  independently by these amounts. The largest of the observed shifts
  for each calorimeter is combined in quadrature.
\item[-]{\bf Particle tracking:}
  The definition of a good charged track has been tightened in this
  analysis in order to minimise possible effects from residual
  tracking distortions primarily in the forward regions of the
  detector. Corrections for the distortions are determined by studying
  Z$\rightarrow\mu^{+}\mu^{-}$ events, and possible remaining
  distortions have been estimated using Bhabha events. The systematic
  uncertainty related to tracking is assessed by applying the
  corrections and adding the remaining distortions independently to Monte
  Carlo event samples and repeating the analysis~\cite{ref-wmass-183}.
\item[-]{\bf Jet energy corrections:}
  Detailed comparisons of reconstructed jets in Monte Carlo and data are
  used to parametrise small corrections to Monte Carlo jet
  energies as function of the jet polar angle to the beam
  axis~\cite{ref-wmass-183}. In order to evaluate the effect from the
  uncertainty in the Monte Carlo jet energy corrections, two alternative
  correction functions, corresponding to $\pm 1\sigma$ errors of the
  discrepancies, are used. The largest shift with respect to the
  nominal correction is taken as the systematic error~\cite{ref-wmass-183}.
\item[-]{\bf Higher order terms:}
  The effect from missing higher order terms, ${\cal O}(\alpha^3)$, in
the simulation of initial state radiation in the {\tt KORALW}
generator, is assessed following the procedure described
  in Ref.~\cite{ref-wmass-183}. In summary, the error on the couplings is
  determined by comparing fits of Monte Carlo samples with events
  weighted to ${\cal O}(\alpha)/{\cal O}(\alpha^2)$ with fits
to the corresponding unweighted samples, which have been generated in
the second order leading-log approximation. Recently, new improved
calculations with the Double Pole Approximation~\cite{ref-DPA} have
appeared. The improved CC03 cross section and angular distributions
predicted by two independent Monte Carlo
programs, {\tt RacoonWW}~\cite{ref-racoon} and {\tt
YFSWW}~\cite{ref-yfsww}, could introduce small changes
on the couplings. For the time being these effects have not been included.
\end{itemize}
\begin{sidewaystable}[p]
\vspace{3mm}
\begin{center}
\renewcommand{\arraycolsep}{1.35}
\small
\begin{tabular}{|l||c|c|c|c|c||c|c|c|c|c||c|c|c|c|c|} \hline
 & \multicolumn{5}{c||}{$\dgz$} &\multicolumn{5}{c||}{$\dkg$} &\multicolumn{5}{c|}{$\lgam$} \\ \hline
Source & 
 \evqq & \muvqq & \tauvqq &  \qqqq & \lvlv & \evqq & \muvqq & \tauvqq & \qqqq & \lvlv & \evqq & \muvqq & \tauvqq & \qqqq & \lvlv \\ \hline \hline
Correlated errors         & \multicolumn{15}{|c|}{ } \\ \hline
Fragmentation             & 0.01 & 0.01 & 0.09& 0.02 &-   & 0.06 & 0.04 & 0.20& 0.05&  - & 0.01 & - & 0.15& 0.04 & -  \\ 
\ww\ cross section        &    - & 0.02& 0.10& 0.02&0.03& - & 0.15& 0.09& 0.04&0.07& - & 0.02& -   & 0.03& -  \\
Luminosity                & - & - & 0.05& - &0.01& - & 0.03 & 0.04& 0.03 &0.03& - & - & - & 0.01 &0.01\\ 
LEP energy                & - & - & - & - & - & - & - & -  & 0.01 & - & - & - & - & - & -  \\ 
W mass                    & - & - & 0.03& - &0.02& 0.02 & - & 0.17& 0.02&0.10& - & - & 0.06& - &0.03\\
Calorimeter scale         & - & - & 0.12& - &0.03& 0.03 & - & 0.12& - &0.04& 0.01 & - & 0.03& - &0.03\\
Tracking                  & - & - & -   & &0.04& - & - & -   & - &0.06& - & - & -   & - &0.01\\ 
Jet corrections           & - & - & 0.01& - & -  & - & - & 0.01& 0.02 & -  & - & - & -   & - & -  \\
Higher order terms        & - & - & 0.01& - & - & - & - & 0.01& - &0.01& - & - & 0.14& - & -  \\ \hline
Uncorrelated errors       & \multicolumn{15}{|c|}{ } \\ \hline
Bose-Einstein correlations& - & - & - & 0.01 & - & - & - & - & 0.02 & - & - & - & - & 0.01 & -  \\
Colour reconnection       & - & - & - & - & - & - & - & - & 0.02 & - & - & - & - & 0.01 & -  \\  
Background estimation     & - & - & - & 0.01 & - & - & - & - & 0.02 & - & - & - & - & 0.01 & - \\ 
Monte Carlo statistics    & - & - & 0.07& &0.24& - & - & 0.19& & -  & - & - & 0.17& - &0.06\\ \hline
Total                     & 0.01 & 0.02 & 0.20 & 0.03 & 0.25 & 0.07 & 0.16 & 0.36 & 0.08 & 0.15 & 0.01 & 0.02 & 0.27 & 0.05 & 0.07 \\ \hline
\end{tabular}
\normalsize
\renewcommand{\arraycolsep}{1.0}
\end{center}
\caption[]
{\label{tab-systematics-189} {\footnotesize Summary of systematic
  uncertainties for the couplings $\dgz$, $\dkg$ and
$\lgam$. A description of the different sources is given in the
text. Systematic uncertainties below 0.005 are indicated by a
dash.}}
\end{sidewaystable}
\begin{sidewaystable}[p]
\vspace{3mm}
\begin{center}
\renewcommand{\arraycolsep}{0.95}
\small
\begin{tabular}{|l||c|c|c|c|c|c|c|c|c|c|c|c|c|c|c|c|} \hline
 & \multicolumn{8}{c|}{Real} &\multicolumn{8}{c|}{Imaginary} \\ \hline
Source & $\tilde{\kappa}_\gamma$ & $\tilde{\lambda}_\gamma$ 
       & $\tilde{\kappa}_Z$ & $\tilde{\lambda}_Z$ 
       & $g^\gamma_4$ & $g^\gamma_5$ 
       & $g^Z_4$ & $g^Z_5$ 
       & $\tilde{\kappa}_\gamma$ & $\tilde{\lambda}_\gamma$ 
       & $\tilde{\kappa}_Z$ & $\tilde{\lambda}_Z$ 
       & $g^\gamma_4$ & $g^\gamma_5$ 
       & $g^Z_4$ & $g^Z_5$ \\ \hline \hline
 Correlated errors       & \multicolumn{16}{|c|}{ } \\ \hline
 Fragmentation           &0.04 &0.03 &0.01 & -   &0.04 &0.01 &0.03 &0.05 &0.02 &0.02 &0.01 &0.01 &0.01 &0.07 & -   &0.02 \\
 \ww\ cross section      &0.02 &0.02 &0.01 &0.01 & -   &0.02 &  -  & -   &0.01 &0.01 &0.01 & -   &0.04 &0.02 &0.02 &0.01 \\
 Luminosity              &0.01 &0.01 & -   & -   &0.01 &0.02 & -   & -   & -   & -   & -   & -   &0.01 &0.02 &0.01 &0.01 \\
 LEP energy              &0.01 & -   & -   & -   &0.02 &0.04 &0.01 &0.03 &0.01 &0.01 & -   & -   &0.01 &0.02 & -   &0.02 \\
 W mass                  & -   & -   & -   & -   & -   &0.02 &  -  & -   & -   & -   & -   & -   &0.01 &0.01 &0.01 & -   \\
 Calorimeter scale       &0.01 &0.01 &0.01 &0.01 & -   &0.04 & -   &0.03 &0.01 &0.01 & -   & -   &0.01 &0.01 & -   &0.01 \\
 Tracking                &0.02 &0.01 &0.01 &0.01 & -   &0.02 & -   & -   &0.01 &0.01 & -   & -   &0.01 &0.02 &0.01 & -   \\
 Jet corrections         & -   & -   & -   & -   &0.01 &0.01 &0.01 &0.01 & -   & -   & -   & -   & -   &0.01 & -   &0.01 \\
 Higher order terms      & -   & -   &  -  &  -  &  -  & -   &  -  & -   &  -  &  -  &  -  &  -  &  -  & -   &  -  &  -  \\ \hline
 Uncorrelated errors     & \multicolumn{16}{|c|}{ } \\ \hline
 Monte Carlo statistics  &0.02 &0.02 &0.02 &0.01 &0.04 &0.04 &0.03 &0.02 &0.01 &0.01 &0.01 &0.01 &0.03 &0.05 &0.02 &0.03 \\
 Background estimation   & -   & -   &  -  &  -  & -   & -   & -   & -   &  -  &  -  &  -  &  -  & -   &  -  &  -  &  -  \\ \hline
 Total                   &0.06 &0.05 &0.02 &0.02 &0.06 &0.08 &0.04 &0.07 &0.03 &0.02 &0.02 &0.01 &0.05 &0.09 &0.04 &0.04 \\
\hline
\end{tabular}
\normalsize
\renewcommand{\arraycolsep}{1.0}
\end{center}
\caption[]
{\label{tab-systematics-189-cp} {\footnotesize Summary of systematic
  uncertainties for the combined \evqq\ and \muvqq\ channels for C- or
P-violating couplings. A description of the different sources is given
in the text. Systematic uncertainties below 0.005 are indicated by a
dash.}}
\end{sidewaystable}\par
Errors assumed to be uncorrelated between channels include: 
\begin{enumerate}
\item[-]{\bf Bose-Einstein correlations:}
  The effect of Bose-Einstein correlations in the \qqqq\ channel is 
  investigated by repeating the analysis on Monte Carlo events
generated with {\tt KORALW} and fragmented using {\tt JETSET} with
Bose-Einstein correlations for all particles, following the
implementation in {\tt
LUBOEI}~\cite{ref-BE}. The scheme for
restoring four-momentum conservation denoted {\tt
BE$_3$}, which has been tuned to the LEP1 Z data, is
considered~\cite{ref-wmass-189}.
\item[-]{\bf Colour reconnection:}
  The uncertainty arising from possible colour reconnection effects is
  assessed by studying Monte Carlo implementations of different colour
  reconnection scenarios in the parton evolution scheme in {\tt
JETSET}~\cite{ref-colour}.  The analysis is repeated with \qqqq\
events generated with the {\tt EXCALIBUR} generator and
  hadronised with and without colour reconnection in the model
referred to as SK1, as described
  in~\cite{ref-wmass-183}. The systematic error is taken as the difference
  in fitted couplings from samples without colour reconnection and
  with colour reconnection in about 30\%\ of the events.
\item[-] {\bf Background estimation:} The error on the couplings from the 
  uncertainties in the background estimation is evaluated by varying the 
  normalisation of the main background processes. The background from 
  QCD is changed by $\pm 5$\%\ based on comparisons between data and 
  Monte Carlo simulation. The background from $\gamma\gamma$, Zee and ZZ 
  processes is varied by $\pm 30$\%\, $\pm 20$\%\  and $\pm 2$\%,
respectively, to account for the theoretical uncertainty in the
prediction for those processes~\cite{ref-lep2report}.
\item[-]{\bf Monte Carlo statistics:}
  The effect of the Monte Carlo statistics is included in the
  systematic uncertainty.
\item[-]{\bf Jet charge assignment:}
  To investigate the effects from the uncertainties on the jet charge,
  the reconstructed W charge is shifted by 0.01. This number is based
  on comparisons between data and Monte Carlo simulation (Z peak
  data)~\cite{ref-jetcharge}. The effect on the couplings is found to
be negligible.
\end{enumerate}

The systematic uncertainties listed above as fully
correlated between 
channels are also assumed to be fully correlated between years. In
addition, the systematic errors from Bose-Einstein correlations and
colour reconnection are taken to be to fully correlated between years.

For both optimal observable methods, OO$_1$ and OO$_2$, the systematic 
errors have been calculated based on the changes in the mean values of
the respective observables. This is incorporated in the TGC extraction
by including the corresponding covariance matrix for the systematic
uncertainties. By this procedure the systematic uncertainties
are folded with the proper statistical correlations between the
optimal observables and the results of
the fits include both the statistical and systematic errors. The
systematic uncertainties listed in Table~\ref{tab-systematics-189} are
derived from the changes in the optimal observables mean values. They are not
used as such in the analysis but serve only as a representation of the
systematic contributions from the different sources.\par

The systematic uncertainties for the maximum likelihood method are
convoluted into the $\log L$ functions by assuming parabolic behaviour
of the systematic errors around the fitted TGC value.\par
\subsubsection{Final results from W-pair production}
\begin{figure}[H]
\begin{center}
\begin{picture}(350,535)
\put(-80,-10){\mbox{\epsfig{file=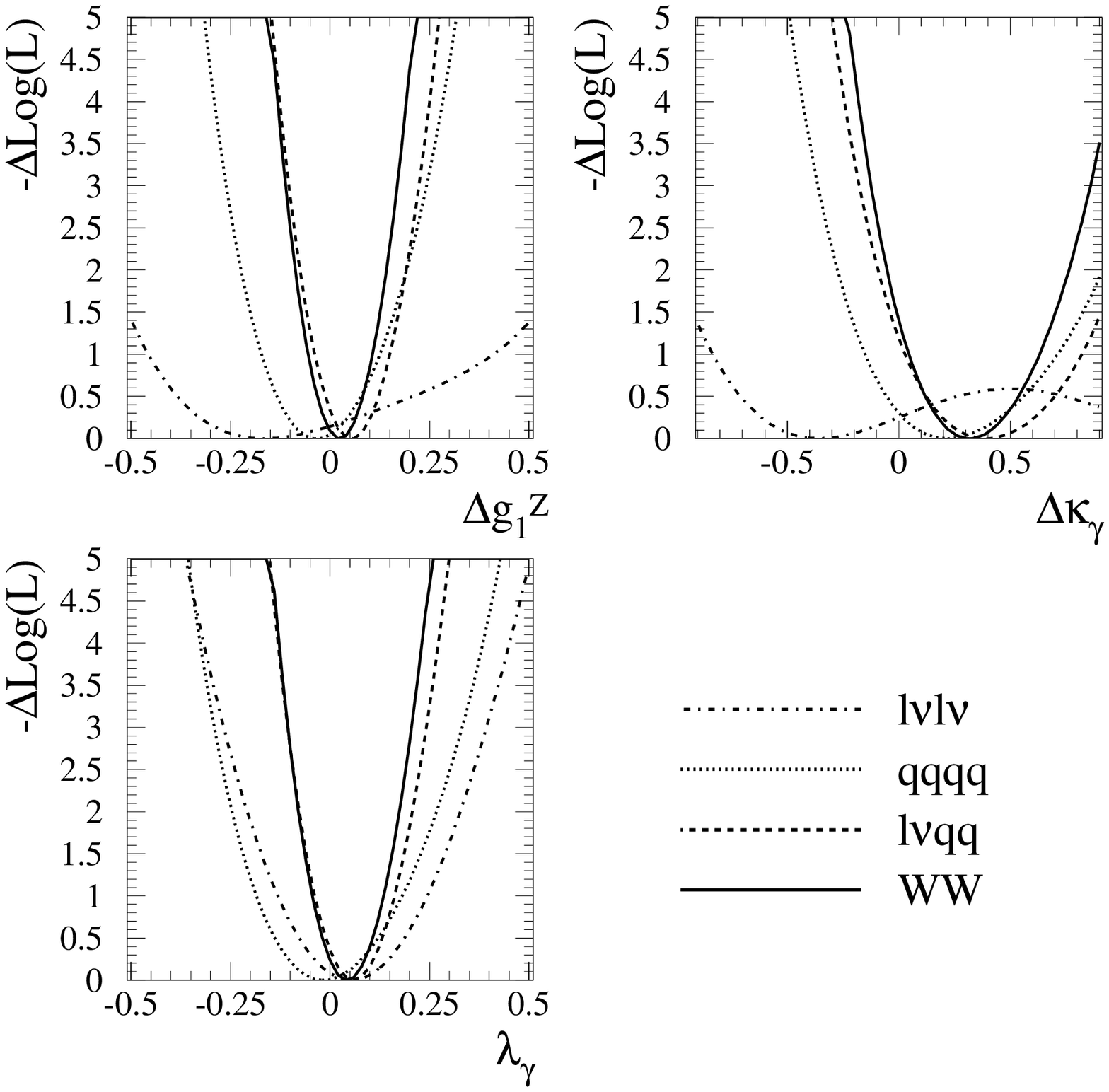,scale=.9}}}
\put(5,405){\Large a)}
\put(245,405){\Large b)}
\put(5,170){\Large c)}
\put(140,482){\bf{\Huge ALEPH}}
\end{picture}
\end{center}
\caption[]
{The combined negative log-likelihood curves from the W-pair analysis
of 183 and 189~\gev\ data for the individual fits in the \lvqq\
(dashed), \qqqq\ (dotted) and \lvlv\ (dashed-dotted) channels for the
three couplings a) $\dgz$, b) $\dkg$ and c) $\lgam$. The curve for
each coupling is obtained while fixing the other couplings to their
Standard Model value. The systematic uncertainties are included. The
combined result for all channels is shown as the solid curve.}
\label{fig-log1d-ww}
\end{figure}

The combined results from all \ww\ decay channels at 183 and 189~\gev\
for the three couplings \dgz, \dkg\ and \lgam, are obtained by combining
the OO$_2$ analysis of the \evqq, \muvqq\ and \qqqq\ final states with the
OO$_1$ analysis of the \tauvqq\ and \lvlv\ final states. The
correlation of the systematic errors between the
different channels and energies are included as described in
Section.~\ref{sec-syst}. The results for \dgz, \dkg\ and 
\lgam, including systematic uncertainties, are
listed in Table~\ref{tab-results-syst}.
\begin{table}\caption[]{\label{tab-results-syst}{\footnotesize
The combined results for 183 and 189~\gev\ for each \ww\ decay channel
for the three couplings \dgz, \dkg\ and \lgam. The error includes the statistical and
systematic uncertainty.}}
\begin{center}
\begin{tabular}{|l||c|c|c|} \hline
 & \multicolumn{3}{|c|}{Coupling} \\ \cline{1-4}
Channel & \dgz & \dkg & \lgam \\ \hline
 \myrule \evqq   & $ ~~0.09_{\T - 0.09}^{\T +0.09}$ & $ ~~0.46_{\T - 0.32}^{\T +0.33}$ & $ ~~0.21_{\T - 0.10}^{\T +0.11}$ \\
 \myrule \muvqq  & $ ~~0.01_{\T - 0.10}^{\T +0.10}$ & $ ~~0.20_{\T - 0.34}^{\T +0.64}$ & $ -0.08_{\T - 0.09}^{\T +0.09}$ \\
 \myrule \tauvqq & $ ~~0.51^{\T + 0.21}_{\T -0.37}$ & $ -0.71^{\T + 0.54}_{\T -0.39}$ & $ ~~0.00^{\T + 0.18}_{\T -0.15}$ \\
 \myrule \qqqq   & $ -0.03_{\T - 0.10}^{\T +0.10}$ & $ ~~0.27_{\T - 0.26}^{\T+ 0.30}$ & $ ~~0.01_{\T - 0.12}^{\T +0.13}$ \\
 \myrule \lvlv   & $ -0.17^{\T + 0.36}_{\T -0.21}$ & $ -0.35^{\T + 0.82}_{\T -0.41}$ & $ ~~0.05^{\T + 0.14}_{\T -0.13}$ \\ \hline
\end{tabular}
\end{center}
\end{table}
The final 68\%\ and 95\%\ combined W-pair result for the three
couplings $\dgz$, $\dkg$ and $\lgam$ is summarised in
Table~\ref{tab-final-ww}. The corresponding $\log L$ curves, including
systematic uncertainties, are shown in Figure~\ref{fig-log1d-ww}.

The  maximum likelihood method is used in the two
semileptonic channels, \evqq\ and \muvqq, to measure the real
and imaginary parts of the C- or P-violating couplings  
$\mathrm{g^V_4}$, $\mathrm{g^V_5}$, $\mathrm{\tilde{\kappa}_{V}}$, and
$\mathrm{\tilde{\lambda}_{V}}$, where V denotes either $\gamma$ or Z.
The combined 183 and 189~\gev\ results, including the systematic
uncertainties, for the C- or P-violating couplings are summarised in
Table~\ref{tab-final-ww}.
\begin{table}[hp]\caption[]
{\label{tab-final-ww} {\footnotesize Combined 183 and
189~\gev\ W-pair results for the three C- and P-conserving couplings,
$\dgz$, $\dkg$ and $\lgam$, and the C- or P-violating
couplings. The error includes the
statistical and systematic uncertainty. The corresponding 95\%\
confidence level intervals are listed in the last column.}}
{\renewcommand{\myrule}{\rule[-4.mm]{0mm}{11mm} }
\begin{center}
\renewcommand{\arraycolsep}{1.35}
\begin{tabular}{|c||c|c|} \hline
 & fit result & 95\%\ confidence limits \\*[-0.5mm] \hline
 \myrule \dgz  &~~$ 0.02^{\T +0.06}_{\T -0.06}$ &  [  -0.09, 0.14 ]  \\*[-0.5mm] \hline
 \myrule \dkg  &~~$ 0.22^{\T +0.21}_{\T -0.20}$ &  [  -0.15, 0.66 ]  \\*[-0.5mm] \hline
 \myrule \lgam &~~$ 0.04^{\T +0.06}_{\T -0.06}$ &  [  -0.08, 0.17 ]  \\*[-0.5mm] \hline\hline
 \myrule $Re(\tilde{\kappa}_\gamma)$ &$  -0.19^{\T + 0.19}_{\T - 0.17}$ &  [  -0.51,  0.18 ]  \\*[-0.5mm] \hline
 \myrule $Re(\tilde{\lambda}_\gamma)$&$   ~~0.17^{\T + 0.14}_{\T - 0.16}$ &  [  -0.15,  0.43 ]  \\*[-0.5mm] \hline
 \myrule $Re(\tilde{\kappa}_Z)$      &$  -0.09^{\T + 0.12}_{\T - 0.11}$ &  [  -0.30,  0.14 ]  \\*[-0.5mm] \hline
 \myrule $Re(\tilde{\lambda}_Z)$     &$   ~~0.07^{\T + 0.09}_{\T - 0.10}$ &  [  -0.12,  0.25 ]  \\*[-0.5mm] \hline
 \myrule $Re(g^\gamma_4)$            &$   ~~0.06^{\T + 0.34}_{\T - 0.35}$ &  [  -0.62,  0.72 ]  \\*[-0.5mm] \hline
 \myrule $Re(g^\gamma_5)$            &$  -0.02^{\T + 0.51}_{\T - 0.51}$ &  [ -1.02,  0.98 ]  \\*[-0.5mm] \hline
 \myrule $Re(g^Z_4)$                 &$   ~~0.07^{\T + 0.23}_{\T - 0.23}$ &  [  -0.38,  0.50 ]  \\*[-0.5mm] \hline
 \myrule $Re(g^Z_5)$                 &$  -0.06^{\T + 0.32}_{\T - 0.31}$ &  [  -0.67,  0.56 ]  \\*[-0.5mm] \hline
 \myrule $Im(\tilde{\kappa}_\gamma)$ &$   ~~0.10^{\T + 0.12}_{\T - 0.12}$ &  [  -0.14,  0.33 ]  \\*[-0.5mm] \hline
 \myrule $Im(\tilde{\lambda}_\gamma)$&$  -0.08^{\T + 0.10}_{\T - 0.10}$ &  [  -0.27,  0.11 ]  \\*[-0.5mm] \hline
 \myrule $Im(\tilde{\kappa}_Z)$      &$   ~~0.03^{\T + 0.08}_{\T - 0.08}$ &  [  -0.13,  0.19 ]  \\*[-0.5mm] \hline
 \myrule $Im(\tilde{\lambda}_Z)$     &$  -0.03^{\T + 0.07}_{\T - 0.07}$ &  [  -0.16,  0.10 ]  \\*[-0.5mm] \hline
 \myrule $Im(g^\gamma_4)$            &$   ~~0.37^{\T + 0.30}_{\T - 0.30}$ &  [  -0.23,  0.95 ]  \\*[-0.5mm] \hline
 \myrule $Im(g^\gamma_5)$            &$  -0.01^{\T + 0.57}_{\T - 0.56}$ &  [ -1.10, 1.10 ]  \\*[-0.5mm] \hline
 \myrule $Im(g^Z_4)$                 &$  ~~ 0.27^{\T + 0.20}_{\T - 0.20}$ &  [  -0.13,  0.65 ]  \\*[-0.5mm] \hline
 \myrule $Im(g^Z_5)$                 &$   ~~0.07^{\T + 0.35}_{\T - 0.35}$ &  [  -0.62,  0.76 ]  \\ \hline
\end{tabular}
\renewcommand{\arraycolsep}{1.00}
\end{center}\renewcommand{\myrule}{\rule[-4.mm]{0mm}{9mm} }}
\end{table}

In all cases described above, each coupling is determined fixing the
other couplings to their Standard Model values. The error intervals
for each coupling are defined as the 68\%\ confidence level intervals
obtained by integration of the likelihood functions, to accommodate
cases with non-parabolic behaviour of the log-likelihood function.
\section{Combined TGC results}
\label{sec-combined}
\begin{figure}[H]
\begin{center}
\begin{picture}(350,520)
\put(-70,-10){\mbox{\epsfig{file=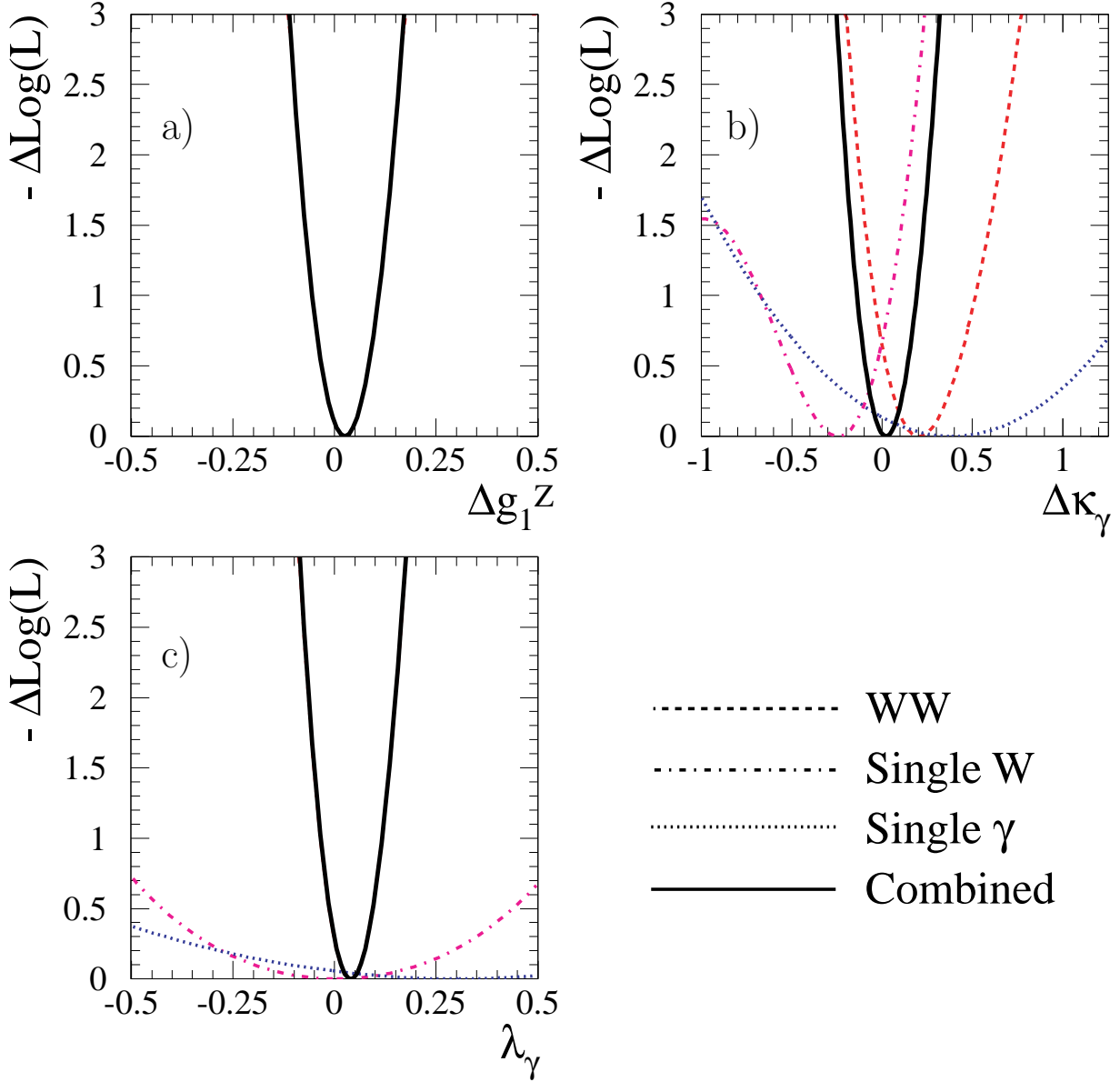,scale=0.9}}}
\put(13,390){\Large a)}
\put(248,390){\Large b)}
\put(13,170){\Large c)}
\put(140,467){\bf{\Huge ALEPH}}
\end{picture}
\end{center}
\caption[]
{
The negative log-likelihood curves for the combined fits using
single-$\gamma$ (dotted), single-W (dashed-dotted) and W-pair
(dashed) production at energies up to 189~\gev\ for the three
couplings a) $\dgz$, b) $\dkg$ and c) $\lgam$. The curve for each
coupling is obtained while fixing the other couplings to their
Standard Model value. The systematic uncertainties are included. The
combined result is shown as the solid curve.}
\label{fig-log1dfinal}
\end{figure}

The measurements from single-$\gamma$, single W and WW
production, are combined with previous ALEPH results from \ww\
production at 172 GeV~\cite{ref-wwtgc-172}, single-W production at 183
\gev~\cite{ref-singlew-183} and single-$\gamma$ production at
183~\cite{ref-singleg-183}. The combined results are listed in
Table~\ref{tab-final-all}.
\begin{table}[t]
\caption[]{ \label{tab-final-all}{\footnotesize
Combined results for $\dgz$, $\dkg$ and $\lgam$ from \ww\ production
at 172-189~GeV, single-$\gamma$ and single-W production at
161-189~\gev. The errors include systematic
uncertainties. The corresponding 95\%\ confidence level intervals are listed
in the last column.}}
\begin{center}
\begin{tabular}{|l||c|c|} \hline
Coupling & fit result & 95\%\ confidence limits \\ \hline
\myrule $\dgz$  
&$~~0.023^{\T +0.059}_{\T -0.055}$ & $ [ -0.087 , 0.141 ]$\\ \hline
\myrule $\dkg$  
&$~~0.022^{\T +0.119}_{\T -0.115}$ & $ [ -0.200 , 0.258 ]$\\ \hline
\myrule $\lgam$ 
&$~~0.040^{\T +0.054}_{\T -0.052}$ & $ [ -0.062 ,0.147 ]$\\ \hline
\end{tabular}
\end{center}
\end{table}
In Figure~\ref{fig-log1dfinal} the corresponding one-parameter $\log
L$ curves are shown.\par

To study the full correlation  between the parameters, two- and 
three-parameter fits, where two or all three couplings are allowed to
vary, are also presented. The fits use the combined information from
W-pair production, single-W production and single-$\gamma$ production
at 183-189 \gev.\par

\begin{figure}[H]
\begin{center}
\begin{picture}(350,530)
\put(-75,-10){\mbox{\epsfig{file=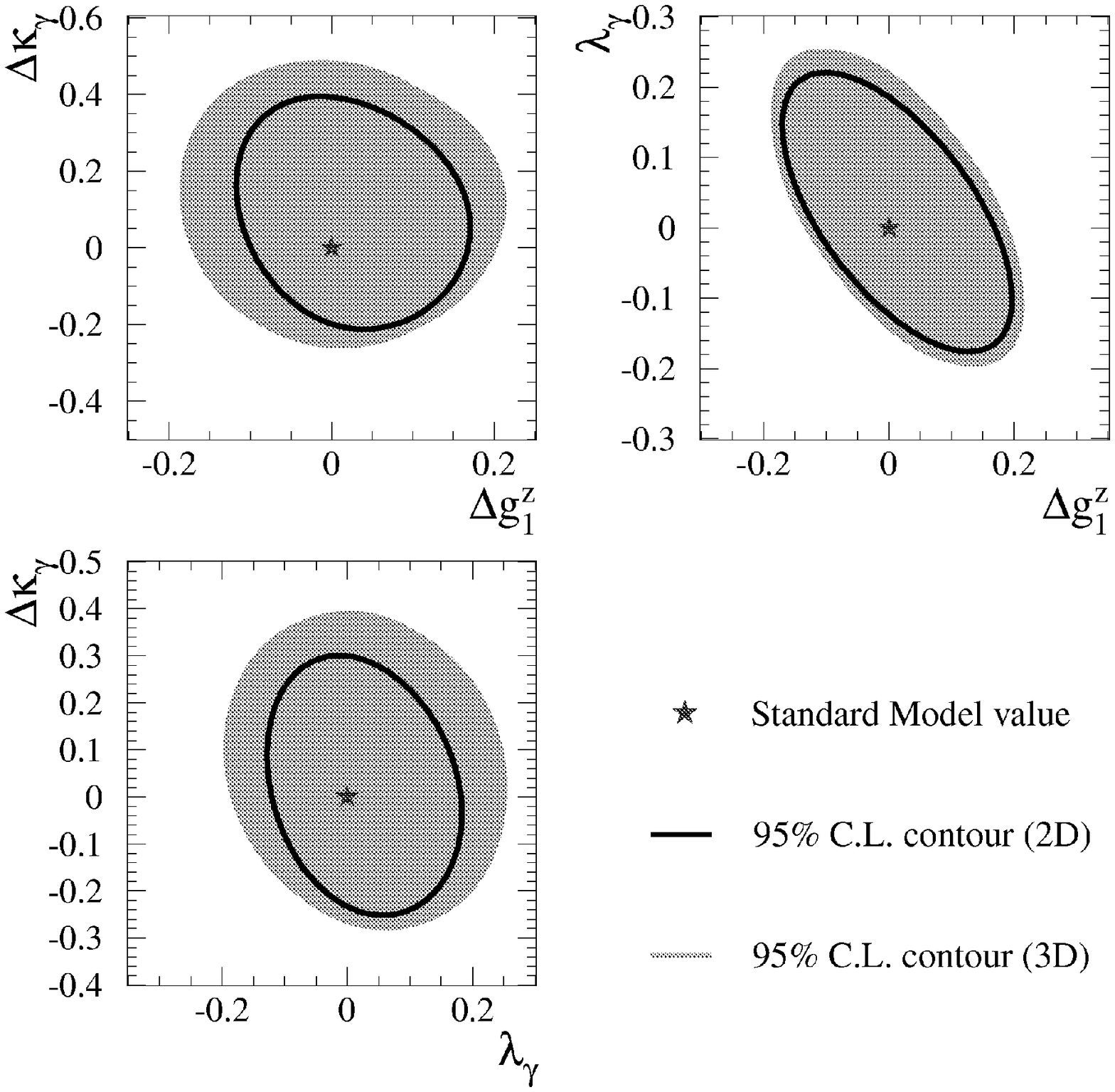,bbllx=22,bburx=590,bblly=112,bbury=680,scale=0.9}}}
\put(10,410){\Large a)}
\put(250,410){\Large b)}
\put(10,170){\Large c)}
\put(140,477){\bf{\Huge ALEPH}}
\end{picture}
\end{center}
\caption[]
{\label{fig-contour} {Multi-parameter fits using the combined
data from single-$\gamma$, single-W and W-pair production at energies
up to 189~\gev. The two-dimensional 95\%\ confidence level contours
for the three pairs of couplings, a) $(\dgz$, $\dkg)$, b) $(\dgz$,
$\lgam)$ and c) $(\dkg$, $\lgam)$. The solid lines show the 95\%\
confidence level contours of the two-parameter fit. The shaded area is
a projection onto the two-dimensional plane of the three-dimensional
envelope of the 95\%\ confidence level volume. The Standard Model point is represented by a star.}}
\end{figure}

For the three parameter fit the results and the
errors
computed from a
variation from the minimum of the $\log L$ functions of 0.5, are
summarised in Table~\ref{tab-final-all-multi} including the systematic
uncertainties. The correlation matrix of the three-parameter fit is
also given in Table~\ref{tab-final-all-multi}. This correlation matrix
is evaluated at the local minimum, and the correlations vary
substantially depending on the exact value of the minimum.
\begin{table}[t]
\caption[]{\label{tab-final-all-multi} {\footnotesize Result of a
three-parameter fit for $\dgz$, $\dkg$ and $\lgam$ using the combined
information from W-pair production at 172-189~GeV, single-$\gamma$ and
single-W production at
161-189~\gev. The statistical and systematic uncertainties are
combined in a 68\%\ one-dimensional error. The corresponding
correlations are given in the last column.}}
\begin{center}
\begin{tabular}{|l||c||ccc|} \hline
 & & \multicolumn{3}{|c|}{Correlation} \\ \cline{3-5}
Coupling & fit result & $\dgz$ & $\dkg$ & $\lgam$ \\ \hline
\myrule $\dgz$ &$~~0.013^{\T +0.066}_{\T -0.068}$ &  1.0  & -0.1 & -0.6 \\ \cline{1-2}
\myrule $\dkg$ &$~~0.043^{\T +0.110}_{\T -0.110}$ &  &  1.0  & -0.1 \\ \cline{1-2}
\myrule $\lgam$&$~~0.023^{\T +0.074}_{\T -0.077}$ &  &  &  1.0  \\ \hline
\end{tabular}
\end{center}
\end{table}
The projections onto the two dimensional plane of the three dimensional
envelope of the 95\%\ confidence level volume, representing the
integration of the confidence over the corresponding third coupling,
are shown in Figure~\ref{fig-contour}. The 95\%\
confidence limits of the respective 2-parameter fits of the three
pairs of couplings $(\dgz$, $\dkg)$,  $(\dgz$, $\lgam)$ and $(\dkg$,
$\lgam)$ are shown as full lines. The systematic
uncertainties are included in the limits shown. No deviations from
the Standard Model expectations are observed.\par
\section{Summary and conclusions}
\label{sec-conclusion}
The triple gauge-boson couplings have been measured using W-pair
events at 183 and 189~\gev, single-W production at 189 \gev\ and
single-$\gamma$ production at 189 \gev. Combining with previous
ALEPH results from \ww\ production at 172 \gev, single-W production
and single-$\gamma$ production at 183~\gev, the
three couplings \dgz, \dkg\ and \lgam\ have been measured individually,
assuming the two other couplings to be fixed at their Standard
Model value. The results are
\begin{center}
\begin{tabular}{lcl}
\myrule $\dgz$ &=&$~~0.023^{\T +0.059}_{\T -0.055}$ \\ 
\myrule $\dkg$ &=&$~~0.022^{\T +0.119}_{\T -0.115}$ \\ 
\myrule $\lgam$&=&$~~0.040^{\T +0.054}_{\T -0.052}$~, \\ 
\end{tabular}
\end{center}
where the error includes systematic uncertainties. The corresponding 95\%\
confidence level limits,
\begin{center}
\begin{tabular}{rcccl}
\myrule &-0.087 $<$ & $\dgz$ & $<$ &0.141, \\
\myrule &-0.200 $<$ & $\dkg$ & $<$ &0.258, \\
\myrule &-0.062 $<$ & $\lgam$ & $<$ &0.147, \\ 
\end{tabular}
\end{center}
are in good agreement with the Standard Model
expectation. Multi-parameter fits, where two or all three couplings
are allowed to vary show also good agreement with the Standard
Model.\par

In addition, semileptonic W-pair events were used to set
limits on the C- or P-violating couplings $\mathrm{g^V_4}$,
$\mathrm{g^V_5}$, $\mathrm{\tilde{\kappa}_{V}}$, and
$\mathrm{\tilde{\lambda}_{V}}$, where V denotes either $\gamma$ or
Z. No deviations from the Standard Model expectations are observed.
\section*{Acknowledgements}
\label{sec-acknow}
It is a pleasure to congratulate our colleagues from the CERN accelerator
divisions for the highly successful operation of LEP at high energies. We
are indebted to the engineers and technicians in all our institutions for
the contributions to the excellent performance of ALEPH. Those of us from
non-member countries thank CERN for its hospitality.

\newpage
\appendix
\section{Leptonic Neural Network Input Variables}
\label{sec-nnvarlep}
The Neural Network (NN) calculates an approximation of the
multidimensional probability density function in the following 13
input variables, for signal and backgrounds. The NN
is applied, after preselection, to events with at least two opposite
charged tracks with momentum -- after bremsstrahlung correction -- in
excess of 15~GeV.  The NN uses variables related to the lepton
candidates, to the missing momentum, global event variables, and WW
kinematics. They are listed here together with their relative
discriminating power, namely the statistical correlation with the
neural network output:

\begin{itemize}
 \item missing mass squared (13.7\%);
 \item missing transverse momentum (11.5\%);
 \item angle between the two most energetic tracks (9.6\%);
 \item energy of the second most energetic track, (8.7\%);
 \item total energy found in a $12^{\mathrm{o}}$ cone around the beam axis (8.2\%);
 \item number of identified leptons with an energy greater than 15~\gev (8\%);
 \item missing transverse momentum with respect to the plane defined
 by the beam axis and the 3D-thrust axis (7.8\%);
 \item energy of the most energetic track (7.4\%);
 \item invariant mass of the two most energetic tracks (6.6\%);
 \item missing longitudinal momentum (6.3\%);
 \item scalar sum of the transverse components of the two most
 energetic tracks with respect to a 2D-thrust axis, built from the
 projection of the track momenta on the transverse
 plane (5.4\%);
 \item number of isolated neutral clusters with energy more than 4 GeV
   outside a cone of $10^{\mathrm{o}}$ around each of  
the two most energetic tracks and forming an invariant mass with each of 
them of more than 2 GeV (4.1\%);
 \item cosine of the angle between the most energetic track and the
   axis perpendicular to the plane defined by the second most
   energetic track and the z-axis (2.6\%).
\end{itemize}
%
%
%

%
%
%
%
\newpage
%
%
\newpage
%
%
\newpage
\newpage
%
%
\newpage
\vspace{5.cm}
\newpage

\newpage
%
%
%
%
%
\mbox{ }
\vspace{3.0cm}
%
%
\mbox{ }
\vspace{3.0cm}

\newpage
\begin{thebibliography}{99}
\bibitem{ref-wwtgc-172} ALEPH Collaboration, {\em Measurement of Triple
  Gauge-Boson Couplings at 172~\gev}, \pl{B422}{1998}{369}.
\bibitem{ref-lepwwtgc} DELPHI Collaboration, {\em Measurement and
interpretation of the W-pair cross-section in $e^+ e^-$ interactions
at 161~GeV}, \pl{B397}{1997}{158};\\
%
DELPHI Collaboration, {\em Measurement of Trilinear Gauge Couplings in
$e^+ e^-$ Collisions at 161~GeV and 172~GeV}, \pl{B423}{1998}{194};\\
%
DELPHI Collaboration, {\em Measurements of the Trilinear Gauge Boson
Couplings WWV (V $\equiv \gamma$, Z) in $e^+ e^-$ Collisions at 183~GeV},
\pl{B459}{1999}{382};\\
%
L3 Collaboration, {\em Pair-Production of W Bosons in $e^+ e^-$
Interactions at $\sqrt{s}=$161~GeV}, \pl{B398}{1997}{223};\\
%
L3 Collaboration, {\em Measurements of Mass, Width and Gauge Couplings of
the W Boson at LEP}, \pl{B413}{1997}{176};\\
%
L3 Collaboration, {\em Measurement of Triple-Gauge-Boson Couplings of the
W Boson at LEP}, \pl{B467}{1999}{171};\\
%
OPAL Collaboration, {\em Measurement of the Triple Gauge Boson Coupling
$\alpha_{W \phi}$ from $W^{+}W^{-}$ Production in $e^+e^-$ Collisions at
$\sqrt{s}=$161~GeV}, \pl{B397}{1997}{147};\\
%
OPAL Collaboration, {\em Measurement of triple Gauge Boson Couplings from
$W^{+}W^{-}$ Production at $\sqrt{s}=$172~GeV}, \epj{C2}{1998}{597};\\
%
OPAL Collaboration, {\em $W^{+}W^{-}$ production and triple gauge boson
couplings at LEP energies up to 183~GeV}, \epj{C8}{1999}{191};\\
%
OPAL Collaboration, {\em Measurement of W Boson Polarisations and
CP-violating Triple Gauge Couplings from $W^{+}W^{-}$ Production at
LEP}, CERN-EP-2000-113 (submitted to Eur.\ Phys.\ J. C);\\
%
OPAL Collaboration, {\em Measurement of triple gauge boson couplings from
$W^{+}W^{-}$ production at LEP energies up to 189 GeV},
CERN-EP-2000-114 (submitted to Eur.\ Phys.\ J. C).
\bibitem{ref-singlew-183} ALEPH Collaboration, {\em A study of single W
  Production in \ee\ collisions at $\sqs =$161-183~\gev},
\pl{B462}{1999}{389}.
\bibitem{ref-singleg-183} ALEPH Collaboration, {\em Measurement of triple
  gauge WW$\gamma$ couplings at LEP2 using photonic events},
  \pl{B445}{1998}{239}.
\bibitem{ref-lepevW} L3 Collaboration, {\em Production of Single W Bosons
at LEP}, \pl{B403}{1997}{168};\\
%
L3 Collaboration, {\em Production of Single W Bosons in \ee\
Interactions at 130~GeV $\leq \sqrt{s}\leq$ 183~GeV and Limits on
Anomalous WW$\gamma$ Couplings}, \pl{B436}{1998}{417};\\
%
L3 Collaboration, {\em Production of Single W Bosons at $\sqrt{s}=$189~GeV
and Measurement of WW$\gamma$ Gauge Couplings}, \pl{B487}{2000}{229}.
\bibitem{ref-tgc-TEV} CDF Collaboration, {\em Observation of W$^+$W$^-$
Production in $p\bar{p}$ Collisions at $\sqrt{s} =$1.8~TeV},
\prl{78}{1997}{4536};\\ 
D\O\ Collaboration, {\em Studies of WW and WZ Production and Limits on
Anomalous WW$\gamma$ and WWZ Couplings}, \prev{D60}{1999}{072002}.
\bibitem{ref-hagivara} K.~Hagiwara, R.~D.~Peccei, D.~Zeppenfeld and K.~Hikasa,
\np{B282}{1987}{253}.
\bibitem{ref-bilenky} M.~Bilenky, J.L.~Kneur, F.M.~Renard and
D.~Schildknecht, \np{B409}{1993}{22}.
\bibitem{ref-lep2report} G.~Gounaris,
  J.-L.~Kneur and D.~Zeppenfeld, from {\em Physics at LEP2}, CERN 96-01
  p. 525, editors G.~Altarelli, T.~Sj\"{o}strand and F.~Zwirner.
\bibitem{ref-lowlep} A.~De~R\'{u}jula, M.~B.~Gavela, P.~Hernandez and
  E.~Mass\'{o}, \np{B384}{1992}{3}.
\bibitem{ref-low} J.~Ellison and J.~Wudka, \ar{48}{1998}{33}.
\bibitem{ref-lmtglow} F. Boudjema {\it et al.}, \prev{D43}{1991}{2223}.
\bibitem{ref-aleph} ALEPH Collaboration, {\em ALEPH: A Detector for
  Electron-Positron Annihilations at LEP}, \nim{A 294}{1990}{121}.
\bibitem{ref-perf} ALEPH Collaboration, {\em Performance of the ALEPH
  Detector at LEP}, \nim{A 360}{1995}{481}.
\bibitem{ref-koralz} S.~Jadach, B.F.L.~Ward and Z.~W\c{a}s, \cpc{79}{1994}{503}.
\bibitem{ref-YFS} D.~R.~Yennie, S.~C.~Frautsch and H.~Suura, \ap{13}{1961}{379}.
\bibitem{ref-nunugpv} G.~Montagna {\it et al.}, \np{B541}{1999}{31}.
\bibitem{ref-grc4f}  J.~Fujimoto {\it et al.}, \cpc{100}{1997}{128}.
\bibitem{ref-hagiwara2} K.~Hagiwara {\it et al.}, \np{B365}{1991}{544}.
\bibitem{ref-photos} E.~Barberio {\it et al.}, \cpc{67}{1991}{115},\ \cpc{79}{1994}{291}.
\bibitem{ref-tauola} S.~Jadach {\it et al.}, \cpc{70}{1992}{69},\ \cpc{76}{1993}{361}.
\bibitem{ref-koralw} M.~Skrzypek, S.~Jadach, W.~Placzek and Z.~W\c{a}s,
  \cpc{94}{1996}{216}. 
\bibitem{ref-pythia} T.~Sj\"{o}strand, \cpc{82}{1994}{74}.
\bibitem{ref-phot02} ALEPH Collaboration, {\em An Experimental Study of
  $\gamma\gamma\rightarrow$Hadrons at LEP}, \pl{B313}{1993}{509};\\
J.~A.~M.~Vermasseren in {\em Procedings of the IVth International
Workshop on Gamma-Gamma Interactions}, Eds.\ G.~Cochard and
P.~Kessler, Springer Verlag (1980).
\bibitem{ref-unibab} H.~Anlauf {\it et al.}, \cpc{79}{1994}{466}.
\bibitem{ref-multig-183} ALEPH Collaboration, {\em Single- and multi-photon 
  production in \ee\ collisions at a centre-of-mass energy of 183~\gev},
  \pl{B429}{1998}{201}.
\bibitem{ref-tsukamoto} T.~Tsukamoto and Y.~Kurihara, \pl{B389}{1996}{162}.
\bibitem{ref-wmass-183} ALEPH Collaboration, {\em Measurement of the W mass
  in e$^+$e$^-$ collisions at 183~\gev}, \pl{B453}{1999}{121}.
\bibitem{ref-wmass-189} ALEPH Collaboration, {\em Measurement of the W
Mass and Width in e$^+$e$^-$ collisions at 189~\gev},
CERN-EP/2000-045 (submitted to Eur.\ Phys.\ J. C).
\bibitem{ref-wxsec-161} ALEPH Collaboration, {\em Measurement of the W mass
  in e$^+$e$^-$ collisions at production threshold}, \pl{B401}{1997}{347}.
\bibitem{ref-wxsec-172} ALEPH Collaboration, {\em Measurement of the W-pair
  cross section in e$^+$e$^-$ collisions at 172~\gev}, \pl{B415}{1997}{435}.
\bibitem{ref-wxsec-183} ALEPH Collaboration, {\em Measurement of W-pair
  cross section e$^+$e$^-$ collisions at 183~\gev}, \pl{B453}{1999}{107}.
\bibitem{ref-oo} M.~Diehl and O.~Nachtmann, \zp{C62}{1994}{397}.
\bibitem{ref-oo2} D.K.~Fanourakis, D.~Fassouliotis and S.E.~Tzamarias,
\nim{A412}{1998}{465};\ \nim{A414}{1998}{399}.
\bibitem{ref-fadin} E.~A.~Kuraev and V.~S.~Fadin,
Sov.\ J.\ Nucl.\ Phys.\  41 (1985) 466.
\bibitem{ref-herwig} G.~Marchesini {\it et al.}, \cpc{67}{1992}{465}.
\bibitem{ref-wxsec-189} ALEPH Collaboration, {\em Measurement of W-pair
  production in e$^+$e$^-$ collisions at 189~\gev}, CERN-EP/2000-052
(submitted to Phys.~Lett.~B).
\bibitem{ref-lepenergy} LEP energy working group, {\em Evaluation of the
LEP Centre-of-Mass energy above the WW production threshold},
CERN-EP/98-191 (submitted to Eur.\ Phys.\ J. C);\\
LEP energy working group, {\em Evaluation of the
LEP Centre-of-Mass energy for data taken in 1998}, LEP Energy Working
Group Note 99/01.
\bibitem{ref-wmass-TEV} CDF Collaboration, {\em A measurement of the W
boson mass}, \prev{D52}{1995}{4784};\\
D\O\ Collaboration, {\em Measurement of the W boson mass at the Fermilab
p$\bar{p}$ collider}, \prl{80}{1998}{3008};\\
Y.~K.~Kim, in proceedings of the Lepton-Photon Symposium 1997, Hamburg,
28 July - 1 August 1997.
\bibitem{ref-DPA} W.~Beenakker, F.~A.~Berends and A.~P.~Chapowsky,
\np{B548}{1999}{3}.
\bibitem{ref-racoon} A.~Denner, S.~Dittmaier, M.~Roth and
D.~Wackeroth, BI-TP 99/45, hep-ph/9912261 (1999).
\bibitem{ref-yfsww} S.~Jadach, W.~Placzek, M.~Skrzypek, B.~F.~L.~Ward
and Z.~W\c{a}s, \pl{B417}{1998}{326}; CERN-TH/1999-222, UTHEP-98-0502.
\bibitem{ref-BE} L.~L{\"o}nnblad and T.~Sj{\"o}strand, \epj{C2}{1998}{165}.
\bibitem{ref-colour} T.~Sj{\"o}strand and V.A.~Khoze,
  \zp{C62}{1994}{281}.
\bibitem{ref-jetcharge} ALEPH Collaboration, {\em Determination of
$sin^2 \theta_{eff}$ Using Jet Charge Measurements in
Hadronic Z Decays}, \zp{C71}{1996}{357}.
\end{thebibliography}
\end{document}